\documentclass[a4paper, amsfonts, amssymb,aps, prb,amsmath, preprint,nofootinbib,floatfix,superscriptaddress,longbibliography]{revtex4-2}%showkeys,twoside
\usepackage[english]{babel}
\usepackage[utf8]{inputenc}
\usepackage[colorinlistoftodos, color=green!40, prependcaption]{todonotes}
\usepackage[pdftex, pdftitle={Article}, pdfauthor={Author}]{hyperref} % For hyperlinks in the PDF
\usepackage{soul} % highlighting
\usepackage[normalem]{ulem} % striking
\usepackage{xcolor} % colored text
\begin{document}
\title{Proximity-Induced Skyrmion Stabilization at the Cu$_2$OSeO$_3$/Bi$_2$Se$_3$ Interface}

\author{S.~Mehboodi} 
\affiliation{School of Natural Sciences, Technical University of Munich, 85748 Garching, Germany} \affiliation{Munich Center for Quantum Science and Technology (MCQST), Munich, Germany} \affiliation{Center for Quantum Engineering (ZQE), Technical University of Munich, 85748 Garching, Germany} 
\author{V.~Ukleev} 
\affiliation{Helmholtz-Zentrum Berlin für Materialien und Energie, Berlin, Germany} 
\author{C.~Luo} 
\affiliation{Helmholtz-Zentrum Berlin für Materialien und Energie, Berlin, Germany} 
\author{R.~Abrudan} 
\affiliation{Helmholtz-Zentrum Berlin für Materialien und Energie, Berlin, Germany} 
\author{J.~Xiao} 
\affiliation{Helmholtz-Zentrum Berlin für Materialien und Energie, Berlin, Germany} 
\author{R.~Golnak} 
\affiliation{Helmholtz-Zentrum Berlin für Materialien und Energie, Berlin, Germany} 
\author{F.~Radu} 
\affiliation{Helmholtz-Zentrum Berlin für Materialien und Energie, Berlin, Germany} 
\author{M.~Kronseder} 
\affiliation{Institute for Experimental and Applied Physics, University of Regensburg, 93040 Regensburg, Germany} 
\author{C.~H.~Back} 
\affiliation{School of Natural Sciences, Technical University of Munich, 85748 Garching, Germany} \affiliation{Munich Center for Quantum Science and Technology (MCQST), Munich, Germany} \affiliation{Center for Quantum Engineering (ZQE), Technical University of Munich, 85748 Garching, Germany} 
\author{A.~Aqeel}\email{aisha.aqeel@uni-a.de}
\affiliation{School of Natural Sciences, Technical University of Munich, 85748 Garching, Germany} 
\affiliation{Munich Center for Quantum Science and Technology (MCQST), Munich, Germany} \affiliation{Institute of Physics, University of Augsburg, 86135 Augsburg, Germany}

\date{\today} % Leave empty to omit a date

\begin{abstract}
    We investigate proximity-induced magnetic interactions at the interface between the topological insulator Bi$_2$Se$_3$ and the chiral magnetic insulator Cu$_2$OSeO$_3$, with particular focus on the low-temperature skyrmion phase. Broadband ferromagnetic resonance spectroscopy reveals enhanced stability of noncollinear spin textures in the Cu$_2$OSeO$_3$/Bi$_2$Se$_3$ heterostructure compared with bare Cu$_2$OSeO$_3$. In addition to an extra resonance mode in the tilted conical phase that is absent in bare Cu$_2$OSeO$_3$, field cycling resolves two counterclockwise skyrmion resonance branches separated by approximately 238 MHz, consistent with the coexistence of a bulk skyrmion lattice and an interfacial skyrmion phase stabilized by proximity-induced exchange coupling and enhanced interfacial Dzyaloshinskii–Moriya interactions. The finite frequency separation indicates that the two skyrmion phases occupy distinct magnetic energy landscapes while retaining similar resonance character. Resonant elastic x-ray scattering measurements further confirm that the interfacial skyrmion phase spans a broader magnetic-field range than the bulk phase, demonstrating enhanced stability and ordering of topological spin textures at the interface. These findings establish interface engineering as a promising route for extending the stability regime of skyrmion and tilted-conical phases in topological–magnetic heterostructures.

\end{abstract}


\maketitle

\section{INTRODUCTION} \label{sec:intro}
\begin{figure*}[ht]
    \centering
    \includegraphics[]{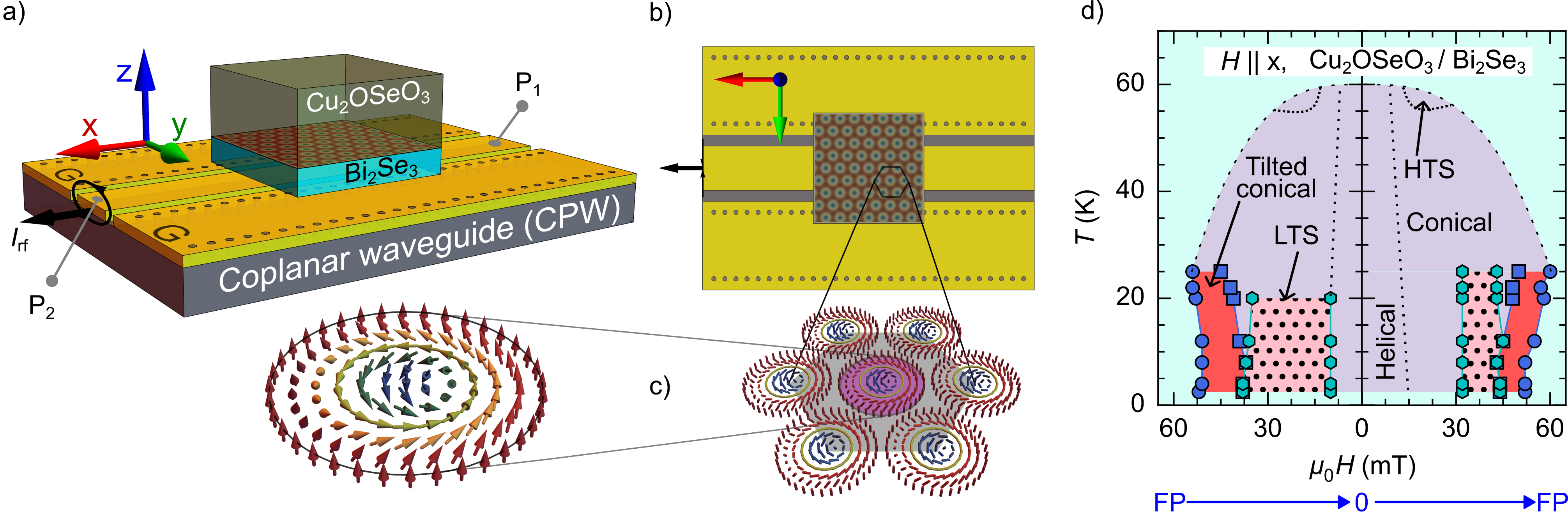}
    \caption{
    (a) Schematic of the Cu$_2$OSeO$_3$/Bi$_2$Se$_3$ (10 QL) heterostructure on a coplanar waveguide (CPW) for broadband magnetic resonance spectroscopy. A radio-frequency current $I_\mathrm{rf}$ is driven by a vector network analyzer while $H$ is applied along $\langle001\rangle$. (b) Top view of the CPW with a hexagonal skyrmion lattice at the interface for $H \parallel {\bf z}$. (c) Enlarged Bloch-type skyrmion; arrow color encodes the out-of-plane component ($+z$ to $-z$). (d) Phase diagram from resonance spectra after zero-field cooling: symbols denote the data measured by magnetic resonance spectroscopy; dashed lines indicate schematic boundaries; arrows show the field-sweep protocol between field-polarized (FP) states.
    }
    \label{cpw}
\end{figure*}

Magnetic skyrmions are topologically nontrivial spin textures that emerge from competing magnetic interactions and have attracted considerable interest for spintronic and magnonic technologies owing to their nanoscale size and efficient manipulation~\cite{koraltan20262026,fert2017magnetic,zhang2015magnetic,nagaosa2013topological,lee2024task}.
In chiral magnets, such textures are stabilized by the Dzyaloshinskii–Moriya interaction (DMI), an antisymmetric exchange that originates from spin–orbit coupling in noncentrosymmetric systems and cants neighboring spins to favor chiral configurations~\cite{camley2023consequences,dzyaloshinsky1958thermodynamic,moriya1960anisotropic}.
A prominent example in this family is Cu$_2$OSeO$_3$~\cite{adams2012long,seki2012observation,muhlbauer2009skyrmion,yu2010real}, an electrical insulator with a large band gap of approximately 2.4~eV~\cite{versteeg2016optically} and ultralow magnetic damping~\cite{stasinopoulos2017low,aqeel2022growth}. Cu$_2$OSeO$_3$ hosts a rich magnetic phase diagram~(Fig.~\ref{cpw}a) with multiple spiral phases including helical, conical, and tilted conical (TC) spiral~\cite{mehboodi2024observation,azhar2022screw,marchiori2024imaging} as well as two independent skyrmion lattice phases, one localized in a small temperature pocket close to the magnetic ordering temperature and another one at lower temperatures~\cite{halder2018thermodynamic,chacon2018observation,bannenberg2019multiple,aqeel2021microwave,lee2021tunable,koraltan20262026}. 
% This rare combination of topologically nontrivial magnetic textures and insulating behavior positions Cu$_2$OSeO$_3$ as a model insulating host of topological spin textures for studying electrically tunable skyrmion dynamics and magnetoelectric coupling
These properties make Cu$_2$OSeO$_3$ a model system for studying the dynamics and stability of topological spin textures~\cite{huang2018situ,white2014electric,okamura2013microwave}.

Recent efforts have been focused on tailoring the stability of topological spin textures through interface engineering, where broken inversion symmetry and strong spin–orbit coupling can modify magnetic interactions~\cite{navabi2019control,fert2017magnetic,gobel2021beyond}. In Cu$_2$OSeO$_3$, interfacial coupling has already been shown to profoundly influence magnetic textures and excitations, enabling the formation of a long-range ordered chiral bobber lattice near the high-temperature skyrmion phase~\cite{ran2021creation} and the hybridization of spin-wave modes with an adjacent ferromagnetic layer~\cite{luthi2023hybrid}. Topological insulators provide a particularly promising route for extending such interfacial control because their spin-polarized surface states can generate proximity-induced exchange interactions and enhance interfacial DMI. Indeed, Bi$_2$Se$_3$-based heterostructures have been shown to exhibit proximity-induced magnetism and strong spin–orbit-driven interfacial effects~\cite{fanchiang2018strongly,lv2018unidirectional,wang2022observation,lv2022large,liu2020changes,wang2016surface,zhu2018proximity,singh2024anisotropic,katmis2016high,he2017tailoring,eremeev2013magnetic}, while skyrmion formation has been reported in several topological-insulator–magnet systems~\cite{yasuda2016geometric,chen2019evidence,li2020topological}. However, how proximity effects arising from a topological-insulator overlayer modify the magnetic phase stability and collective spin dynamics of bulk insulating chiral magnets remains largely unexplored.

% Surface engineering of \CSO single crystal enables the pinning and creation of long-range ordered chiral bobber lattice near the high-temperature skyrmion (HTS) phase pocket, driven by proximity coupling through a multilayer ferromagnetic thin film~\cite{ran2021creation}. Coupling \CSO to an adjacent ferromagnetic layer has likewise been shown to hybridize the spin-wave modes of the two systems at the interface, as resolved by ferromagnetic resonance~\cite{luthi2023hybrid}.

% Topological insulators (TIs) such as Bi$_2$Se$_3$ (\bss) are a natural choice of interfacial partner because their strong spin–orbit coupling, together with the broken inversion symmetry at the interface, can modify the magnetic interactions of an adjacent layer.
% Recent studies on \BS thin films integrated with ferromagnetic and antiferromagnetic layers have revealed proximity-induced magnetism and spin–orbit coupling effects~\cite{fanchiang2018strongly,lv2018unidirectional,wang2022observation,lv2022large,liu2020changes,wang2016surface,zhu2018proximity,singh2024anisotropic,katmis2016high,he2017tailoring,eremeev2013magnetic}, providing a fertile platform to explore the interplay between magnetic order and topological states~\cite{navabi2019control,fert2017magnetic,gobel2021beyond} and to support skyrmion formation in TI/magnetic-insulator systems~\cite{yasuda2016geometric,chen2019evidence,li2020topological}.

In bulk Cu$_2$OSeO$_3$~crystals, the formation of the low-temperature skyrmion (LTS) phase requires overcoming an energy barrier, often necessitating protocols such as magnetic field cycling to stabilize and populate the skyrmion phase~\cite{lee2021tunable,aqeel2021microwave}. In this work, we investigate proximity-induced effects in a Cu$_2$OSeO$_3$/Bi$_2$Se$_3$ heterostructure using broadband ferromagnetic resonance spectroscopy, complemented by resonant elastic X-ray scattering (REXS) and magnetometry. We observe an additional resonance mode in the TC phase and enhanced stability of the LTS phase compared with bare Cu$_2$OSeO$_3$. Resolving the breathing and counterclockwise (CCW) skyrmion modes~\cite{mochizuki2012spin,onose2012observation,schwarze2015universal,koraltan20262026}, we further find that the CCW resonance splits into two branches separated by approximately 238 MHz, consistent with the coexistence of bulk and interfacial skyrmion phases.
Combined with the expanded skyrmion stability region observed by REXS, these results indicate that proximity coupling at the Cu$_2$OSeO$_3$/Bi$_2$Se$_3$ interface modifies the magnetic energy landscape and promotes the stabilization of well-ordered topological spin textures. Our findings establish topological-insulator interfaces as a promising route for controlling the phase stability and collective excitations of skyrmions in insulating chiral magnets.

% Resolving the breathing and counterclockwise (CCW) modes of the skyrmion lattice~\cite{mochizuki2012spin,onose2012observation,schwarze2015universal,koraltan20262026}, we further observe that the CCW mode splits into surface- and bulk-related branches separated by $\sim$238~MHz, corroborated by resonant elastic X-ray scattering. We attribute these effects to a proximity-induced enhancement of the interfacial DMI arising from the strong spin–orbit coupling (SOC) of Bi$_2$Se$_3$ and the broken inversion symmetry at the interface, which also modifies the surrounding spin textures into well-ordered phases. Disentangling the specific contribution of the topological surface states of Bi$_2$Se$_3$ from a more general interfacial spin–orbit effect is an interesting direction for future work. These results identify the Cu$_2$OSeO$_3$/Bi$_2$Se$_3$ interface as a route to stabilize and tune skyrmions in a fully insulating heterostructure.
%%%%%%%%%%%%%%%%%%%%%%%%%%%%%%%%%%%%%%%%

\section{EXPERIMENTAL METHODS} \label{sec:method}

We used broadband magnetic resonance spectroscopy to study magnetization dynamics in chiral magnet/topological insulator heterostructures, focusing on the Cu$_2$OSeO$_3$/Bi$_2$Se$_3$ system. The Cu$_2$OSeO$_3$ single crystal, grown via chemical vapor transport~\cite{aqeel2022growth}, was shaped into a platelet with dimensions of 1.2~mm~\(\times\)~1.2~mm~\(\times\)~0.5~mm and \(\langle\)001\(\rangle\) crystallographic direction normal to the sample surface. Due to the cubic symmetry of Cu$_2$OSeO$_3$, the out-of-plane (${\bf z}\parallel[001]$) and in-plane (${\bf x}\parallel[100]$) field directions are equivalent $\langle001\rangle$-type axes, denoted $\langle001\rangle$ throughout.
10 QL (quintuple layers) of Bi$_2$Se$_3$ were deposited on a well-oriented polished surface by molecular beam epitaxy. The sample was covered by (10nm) Se / (7nm) AlO$_x$ /(2nm) Al to prevent oxidation.

As illustrated in Figure~\ref{cpw}(a), the sample was mounted with the Bi$_2$Se$_3$ facing a coplanar waveguide (CPW). The CPW was connected to the ports P$_1$ and P$_2$ of a vector network analyzer (VNA), which served simultaneously as the rf current source and detector. The VNA provides the complex transmission parameter \(S_{21}\) (i.e., the signal transmitted from P$_1$ to P$_2$) as a function of frequency under a fixed external magnetic field $H$. The
measurements were conducted at a temperature of 5~K, sweeping the magnetic field from a field-polarized (FP) state down to zero, and then again backwards towards the FP state in 1~mT steps.
At low temperatures, Cu$_2$OSeO$_3$ exhibits a range of unconventional noncollinear magnetic phases arising from the interplay between cubic magnetocrystalline anisotropy, exchange interactions, and DMI \cite{chacon2018observation}.
% With decreasing field, the system evolves from the field-polarized state through the conical phase below  \(H_{\text{c2}} \) to the helical phase below  \(H_{\text{c1}} \), while the TC state and the LTS phase, which is stabilized when the magnetic field is applied along the \(\langle\)001\(\rangle\) direction
With decreasing field, the system evolves from the FP state through the conical phase below \(H_{\text{c2}} \) to the helical phase below \(H_{\text{c1}} \) and for \(H \parallel\langle001\rangle\), the TC and LTS phases additionally appear within this range~\cite{chacon2018observation,qian2018new,halder2018thermodynamic}.
All measurements were followed by a zero-field cooling protocol, and magnetic field cycling was applied to overcome metastable energy barriers and facilitate skyrmion nucleation in the low-temperature regime.

To suppress background noise and normalize the spectra to remove loss-related background, we applied a post-processing technique known as the derivative divide method~\cite{maier2018note}, which computes the field derivative of the normalized signal, yielding Re$(\partial S_{21}/\partial H)$ (see Sec.~\ref{app_fit} of the Supplemental Material~\cite{supplemental_material}).
Figure~\ref{cpw}(d) shows the magnetic phase diagram of Cu$_2$OSeO$_3$ obtained from magnetic resonance spectroscopy measurements under an in-plane magnetic field ($H \parallel {\bf x}$). The low-temperature range (2.5-25~K, see Sec.~ \ref{sec:tdependency} of the Supplemental Material~\cite{supplemental_material}) corresponds to the experimentally measured data indicated by solid lines, while the dashed lines represent the schematic phase boundaries extrapolated to higher temperatures.

Resonant elastic X-ray scattering (REXS) measurements were carried out using the ALICE II endstation at the UE51$\_$PGM-1 OPUS undulator beamline, BESSY II (Helmholtz-Zentrum Berlin) in reflection geometry, as shown in Fig.~\ref{fig:Rexs_geo}. The incident soft X-rays were tuned to the Cu $L_3$ absorption edge (photon energy $\approx 932$~eV) to enhance magnetic contrast via resonant scattering from the Cu$^{2+}$ moments. The scattering angle was set to $2\theta \approx 96.5^\circ$, allowing detection of magnetic satellite peaks around the structural Bragg reflection. These magnetic satellites arise from the long-wavelength spin modulations and provide direct access to the helical, skyrmion (LTS and HTS), and TC magnetic phases in reciprocal space.

\section{Results} \label{sec:results}

\begin{figure*}[]
    \centering
    \includegraphics[]{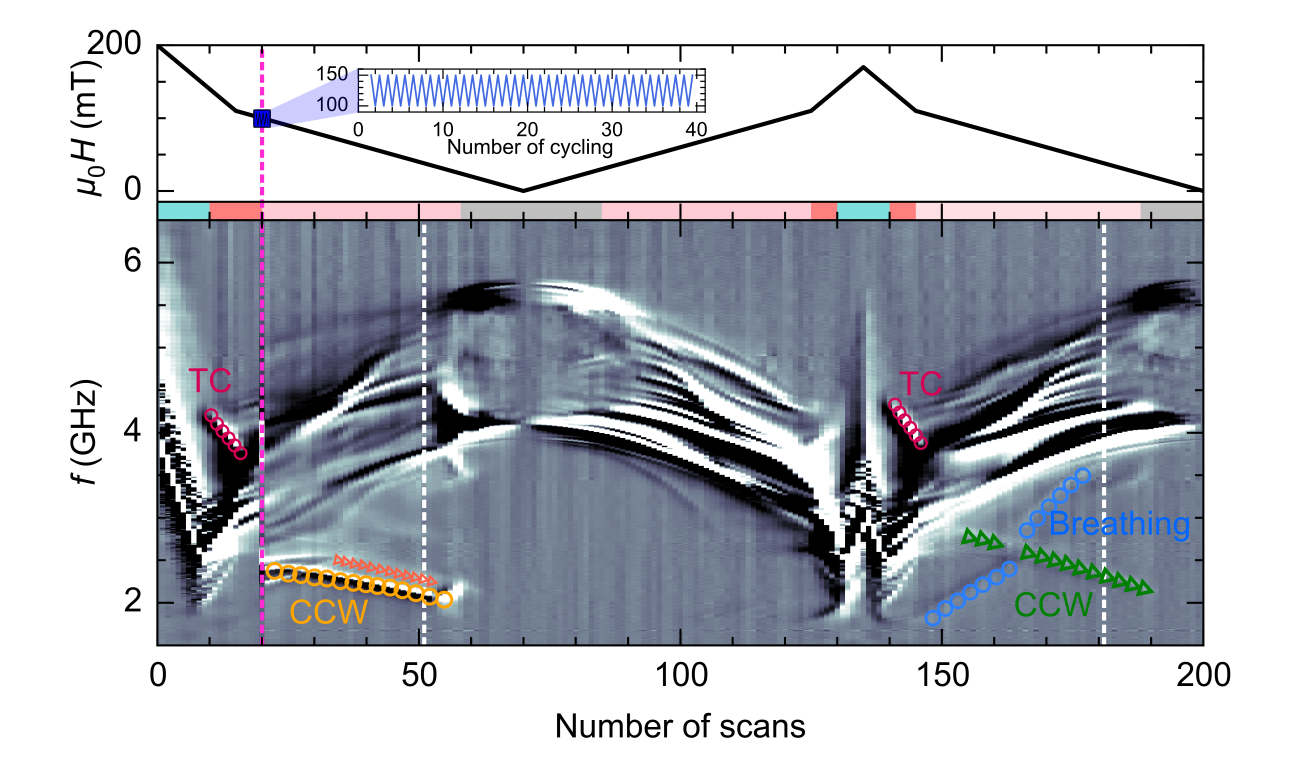}
    \caption{Experimental microwave transmission spectra of Cu$_2$OSeO$_3$/Bi$_2$Se$_3$ at 5~K with \(H\) $\parallel {\bf z}$. The field-sweep protocol is shown in the top panel with insert highlighting a field cycling protocol at that specific field value.
    The top-panel color bar indicates the regions identified from the FMR spectra: FP (cyan), TC/conical (red), skyrmion/conical (pink), and helical (gray).
    Orange and green markers highlight the CCW modes observed with and without field cycling, respectively, while blue and red symbols denote breathing and TC modes.}
    \label{spec22}
\end{figure*}

\subsection*{Ferromagnetic Resonance Spectroscopy}

Figure~\ref{spec22} shows the field-dependent magnetic resonance spectra of the Cu$_2$OSeO$_3$/Bi$_2$Se$_3$ heterostructure for \(H\parallel\langle001\rangle\). 
% by employing a VNA to measure $S_{21}$ across the frequency spectrum from 1.5~GHz to 6.5~GHz for each individual magnetic field scan.
The spectra were obtained by measuring the VNA transmission parameter $S_{21}$ as a function of excitation frequency at a fixed magnetic field. The magnetic field was initially reduced from the FP state at \(\mu_0H=200\)~mT to zero, subsequently increased back to the FP state, and finally reduced to zero again, allowing comparison of the resonance spectra before and after field cycling. 

Between scans 0 and 11, the sample remains in the FP regime and exhibits the uniform Kittel mode together with standing spin-wave resonances arising from the exceptionally low magnetic damping of Cu$_2$OSeO$_3$ at low temperatures~\cite{stasinopoulos2017low} (see Fig.~\ref{ssw-Tdependency} in the Supplemental Material~\cite{supplemental_material}). At scan 20 (indicated by a purple dashed line 
%and the blue marker in the top panel
), 40 field cycles were applied between \(\mu_0H_\text{low}=\)~100~mT and \(\mu_0H_\text{high}=\)~150~mT (inset of Fig.~\ref{spec22}), a protocol known to enhance the population of the LTS phase. The CCW resonance, marked by orange circles, is observed between approximately 2.0 and 2.4~GHz over a field range extending from about 90~mT down to 20~mT. Near the critical field $H_\text{c1}$, the mode exhibits a distinct kink accompanied by a reversal of its field dependence.
%Furthermore, a kink was observed at the tail end of the LTS mode, indicating a sign change of the mode's slope near \(\mu_0H_{c1}\).

After returning the sample to the field-polarized state, the magnetic field was decreased again without additional cycling (scans 135–200). In this measurement branch, the characteristic breathing and CCW skyrmion modes, highlighted by blue and green markers, become clearly resolved. As discussed below, the evolution and subsequent splitting of the CCW resonance provide evidence for distinct bulk and interfacial skyrmion populations in the Cu$_2$OSeO$_3$/Bi$_2$Se$_3$ heterostructures.

Additionally, as the magnetic field is reduced from the field-polarized state toward zero, a distinct resonance mode emerges upon crossing the critical field \(H_{\text{c2}} \) (see Fig.~\ref{app_spec} of the Supplemental Material~\cite{supplemental_material}). This excitation is not observed in the magnetic resonance spectra of bare Cu$_2$OSeO$_3$~\cite{aqeel2021microwave}, indicating that the Bi$_2$Se$_3$ overlayer modifies the magnetic phase behavior near the interface. The mode appears within the field range associated with the TC phase, an intermediate magnetic state reported near \(H_{\text{c2}} \) that is often difficult to resolve in bare Cu$_2$OSeO$_3$ due to phase coexistence and spectral overlap.

The emergence of a well-defined TC resonance suggests that proximity effects at the Cu$_2$OSeO$_3$/Bi$_2$Se$_3$ interface enhance the stability and ordering of the underlying spin texture. A plausible origin is an interfacial modification of the magnetic interactions, including an enhancement of the effective DMI arising from the strong spin–orbit coupling of Bi$_2$Se$_3$ and the broken inversion symmetry at the interface. Such an effect would favor a more coherent magnetic configuration and could account for the appearance of a distinct collective excitation that is absent in bare Cu$_2$OSeO$_3$.

\begin{figure*}[ht]
    \centering
    \includegraphics[]{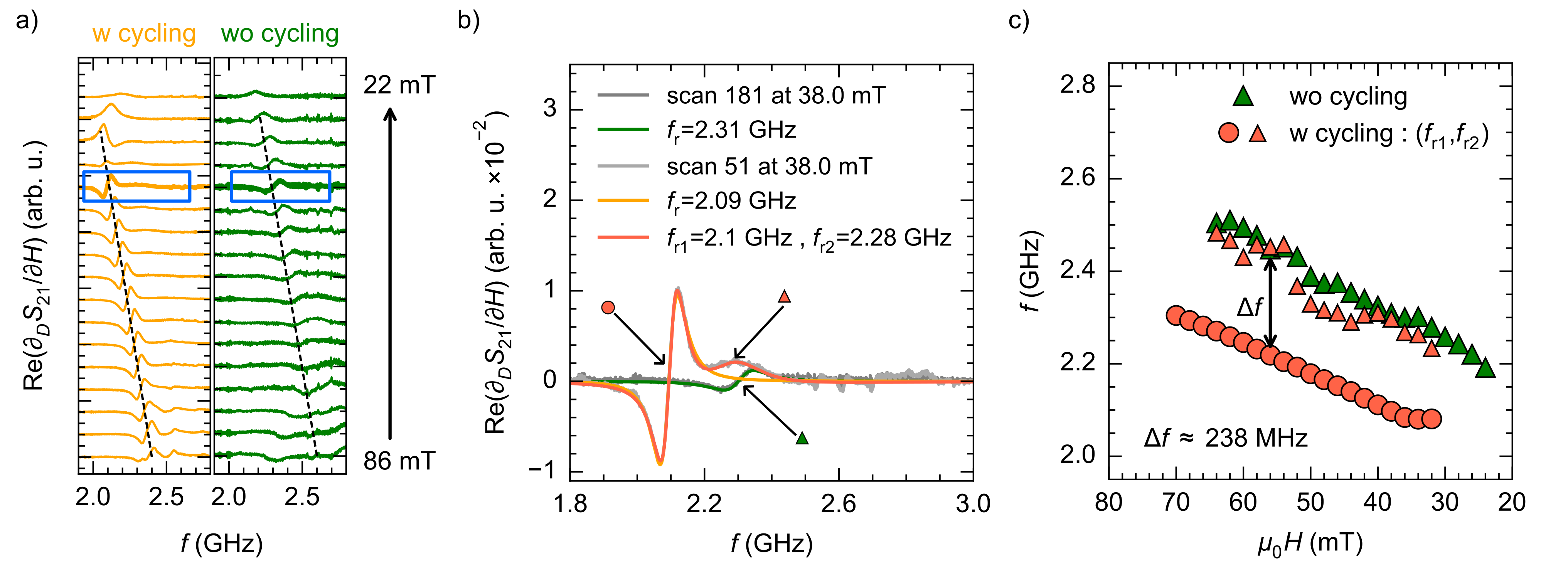}
    \caption{(a,b) Line scans of the resonance spectra highlighting the CCW mode with (orange) and without (green) magnetic field cycling. The line scans are measured at fixed fields as a function of frequency. 
    The dashed lines in (a) highlight the resonance frequency shift as a function of applied magnetic field. (b) shows the zoomed-in view of two resonance line scans (51: without cycling and 181: with field cycling), also highlighted with blue rectangles in (a).
    (c) Comparison of field evolution of the resonance frequencies of the CCW mode obtained without field cycling (green) and with field cycling (orange).}
    \label{all_plot}
\end{figure*}

Figure~\ref{all_plot}(a) presents frequency-dependent spectra measured in the skyrmion-field range (\(22\leq \mu_0H \leq 86\) mT) for both the field-cycled (orange) and non-cycled (green) measurement protocols. The skyrmion mode region is highlighted, and resonance peaks are traced with black dashed lines for visual clarity. The resonance positions are indicated by black dashed lines to guide the eye. While both datasets exhibit the characteristic CCW skyrmion resonance, a pronounced difference is observed after field cycling.

To illustrate this behavior, Fig.~\ref{all_plot}(b) compares spectra recorded at \(\mu_0H=38\) mT, corresponding to the field cut indicated in Fig.~\ref{all_plot}(a)). In the non-cycled measurement, the CCW mode appears as a single resonance centered at \(f_\text{r1}\). After field cycling, however, the resonance acquires a pronounced asymmetric lineshape consisting of a dominant peak at \(f_\text{r1}\) together with a shoulder at the higher frequency \(f_\text{r2}\). Fits to the spectra confirm the presence of two closely spaced resonances in the cycled state. The extracted resonance frequencies are summarized in Fig.~\ref{all_plot}(c). Across the entire skyrmion-field range, the two branches remain separated by up to $\approx$~238 MHz, demonstrating that the field-cycled state cannot be described by a single CCW resonance mode. Instead, the data are consistent with the coexistence of two skyrmion populations experiencing different magnetic environments.

A natural interpretation is that the higher- and lower-frequency branches originate from skyrmions associated with the interfacial and bulk regions of the Cu$_2$OSeO$_3$/Bi$_2$Se$_3$ heterostructure, respectively. In this picture, proximity-induced modifications of the magnetic interactions at the interface, including an enhanced effective DMI arising from strong spin–orbit coupling and broken inversion symmetry, stabilize an interfacial skyrmion phase distinct from the bulk skyrmion lattice. The persistence of the frequency splitting over a broad field range therefore provides evidence that the Bi$_2$Se$_3$ overlayer modifies the skyrmion energy landscape and promotes the stabilization of interfacial topological spin textures.

Using the measured CCW dispersion ($df/d\mu_0H \approx 7$~GHz/T), the $\Delta f \approx 238$~MHz branch separation corresponds to an effective-field difference of $\approx 34$~mT between the bulk and interfacial skyrmion environments (Sec.~\ref {app:field}, Supplemental Material~\cite{supplemental_material}). Ascribing this to an enhanced interfacial DMI implies an interfacial propagation vector larger by a few percent ($\Delta q/q \approx 5\%$, a pitch $\sim$3~nm shorter), which would broaden rather than split the low-temperature skyrmion satellites in REXS, consistent with the slightly enhanced LTS linewidth observed here.

Recently, Fedel \textit{et al.} showed that interfacial DMI in ferrimagnetic-insulator/nonmagnetic-metal heterostructures can extend beyond the immediate atomic interface through spin--orbit-mediated coupling mechanisms~\cite{fedel2025evidence}. Motivated by this result, we consider analogous proximity-induced modifications at the Cu$_2$OSeO$_3$/Bi$_2$Se$_3$ interface. In the present system, the strong spin--orbit coupling of Bi$_2$Se$_3$ and the broken inversion symmetry at the interface may modify the local magnetic free-energy landscape, potentially through interfacial DMI, exchange coupling, anisotropy, and boundary-condition effects. Such interfacial modifications provide a plausible origin for the enhanced stability of noncollinear spin textures observed in our measurements.
Within this picture, the coexistence of bulk and interfacial skyrmion populations offers a plausible explanation for the two closely spaced CCW resonance branches resolved in the field-cycled state. Similar proximity-induced modifications of magnetic interactions have been predicted and observed in topological-insulator/magnet heterostructures~\cite{yasuda2016geometric,chen2019evidence}.

%%%%%%%%%%%%%%%%%%%%%%%%%%%%%%%%%%%%%%%%%%%%%%%%%%

\subsection*{Resonant elastic X-ray scattering}{\label{REXS-manuscript}}

\begin{figure*}
    \centering
    \includegraphics[]{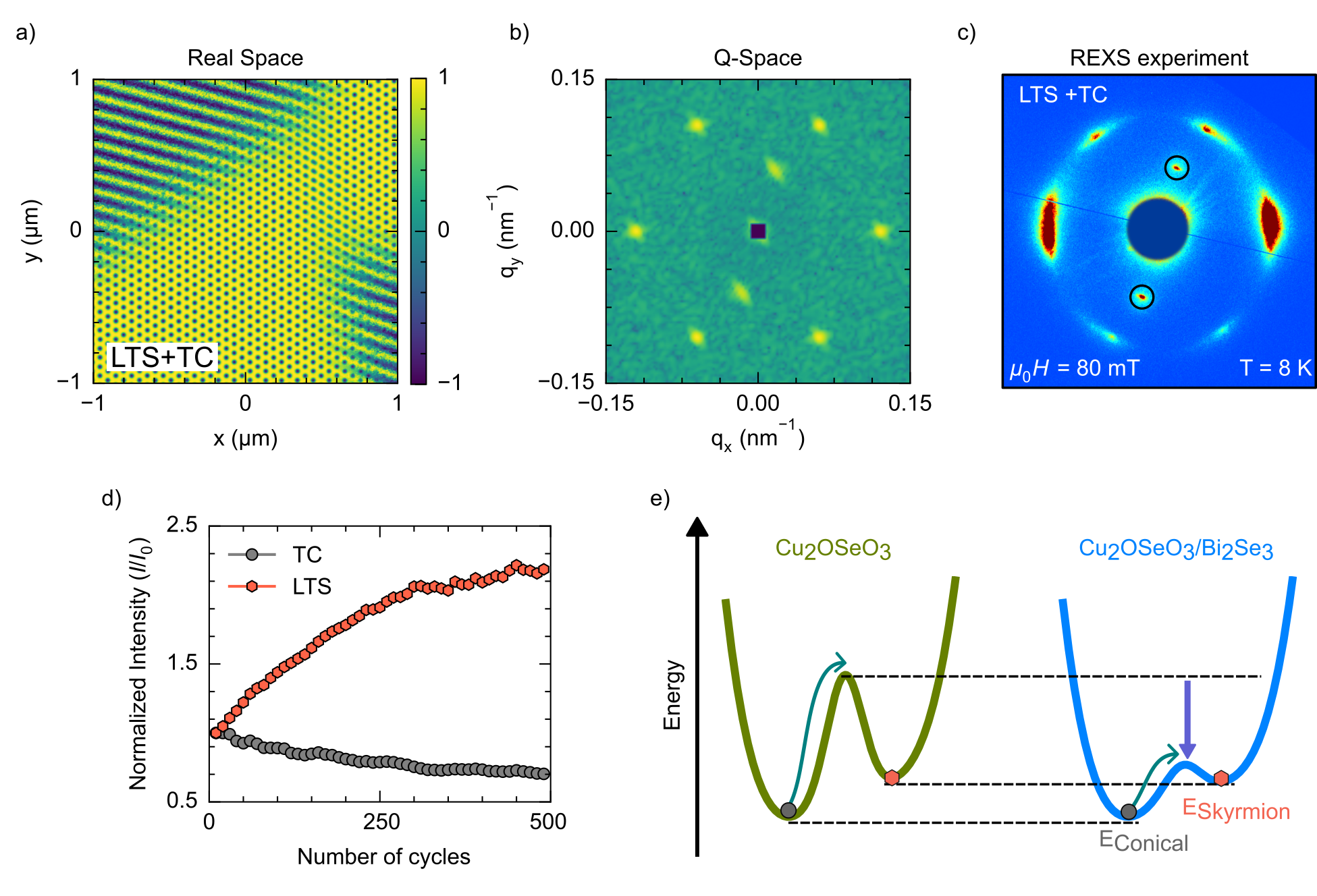}
    \caption{Phenomenological model showing (a) real-space magnetic configurations and (b) synthetic REXS. (c) shows the experimentally observed REXS pattern measured after field cycling.
    (d) Normalized integrated intensity of the magnetic satellites as a function of magnetic-field cycling, showing the progressive enhancement of the LTS signal and suppression of the TC contribution. 
    (e) Schematic energy landscape comparing bare Cu$_2$OSeO$_3$ and Cu$_2$OSeO$_3$/Bi$_2$Se$_3$ samples, illustrating the reduced energy barrier between the conical and skyrmion states in the heterostructure.
    }
    \label{fig:REXS_cycle}
\end{figure*}

We performed REXS measurements on the Cu$_2$OSeO$_3$/Bi$_2$Se$_3$ heterostructure (see Fig.~\ref{fig:Rexs_geo}). In contrast to our FMR results, no clear signature of the LTS phase was observed without applying a magnetic-field cycling protocol. This behavior can be understood from the intrinsic sensitivity of REXS to long-range periodic order: a well-defined diffraction pattern in reciprocal space requires a coherent and spatially ordered spin texture, while the dynamic
response from skyrmions, irrespective of their long‑range spatial order, can be captured in the FMR measurement. In the Cu$_2$OSeO$_3$/Bi$_2$Se$_3$ sample, the absence of an initial LTS signal suggests that small skyrmion clusters may be spatially randomly distributed and stabilized at the interface, preventing the formation of a sufficiently ordered lattice detectable by REXS (see Fig.~\ref{fig:REXS_all}(b) of the Supplemental Material~\cite{supplemental_material}).

Upon magnetic-field cycling (see Fig.~\ref{fig:REXS_cycle}(d)), a clear LTS diffraction signal emerges after approximately ten cycles, characterized by six well-defined magnetic satellites surrounding the forbidden (001) structural Bragg peak. Compared to bare Cu$_2$OSeO$_3$, the LTS phase in the Cu$_2$OSeO$_3$/Bi$_2$Se$_3$ sample appears to be more readily populated, indicating a reduced energy barrier between the spiral and skyrmion local minima in the magnetic energy landscape as shown in Fig.~\ref{fig:REXS_cycle}(e).
In bare Cu$_2$OSeO$_3$, field cycling populates the LTS as a multidomain state with a ring-like intensity distribution and no fixed in-plane orientation; obtaining a single-domain lattice (a well-defined six-fold diffraction pattern) requires an additional step of tilting the field a few degrees away from $\langle001\rangle$ and back~\cite{chacon2018observation}. In our Cu$_2$OSeO$_3$/Bi$_2$Se$_3$ sample, by contrast, field cycling alone already yields the single-domain, six-fold pattern, without this additional field-tilting procedure.

To interpret the measured diffraction pattern, we model the near-surface texture as coexisting LTS and TC order, as shown in Fig.~\ref{fig:REXS_cycle}(a); the corresponding synthetic REXS pattern in Fig.~\ref{fig:REXS_cycle}(b) reproduces the six-fold magnetic satellites observed experimentally in Fig.~\ref{fig:REXS_cycle}(c).
The corresponding synthetic reciprocal-space patterns were obtained from the Fourier intensity of phenomenological real-space magnetic textures, as described in Sec.~\ref{REXS} of the Supplemental Material~\cite{supplemental_material}.
As shown in Fig.~\ref{fig:REXS_cycle}(c), the magnetic satellites associated with the LTS phase are not equally distributed among the six symmetry-related positions. Therefore, full rocking scans were performed in order to integrate over the out-of-plane component ($\text{Q}_z$) and accurately determine the total scattered intensity of each satellite.

Furthermore, the LTS magnetic satellites exhibit a slightly larger linewidth compared to those of the HTS phase (see Fig.~\ref{fig:REXS_all}(d3)). The magnetic coherence length $\xi$ was estimated from the radial peak width according to
$
\xi \propto  \frac{1}{\Delta \text{Q}},
$
where $\Delta Q$ denotes the full width at half maximum of the magnetic satellite. The increased peak broadening observed in the LTS regime, therefore, indicates a reduced coherence length of the single-domain skyrmion lattice. This behavior is consistent with the coexistence of the TC phase, which introduces additional spatial modulation and limits the long-range coherence of the skyrmion lattice. Part of this broadening is also consistent with the small ($\sim$5\%) bulk--interfacial difference in propagation vector inferred from the CCW splitting (see Sec.~\ref {app:field} of the Supplemental Material~\cite{supplemental_material}).

Moreover, a direct comparison between surface-sensitive REXS and bulk SQUID susceptibility measurements reveals distinct magnetic phase stability at the interface and in the bulk. While SQUID measurements readily detect the LTS region, REXS shows that only a few magnetic-field cycles are required to stabilize a well-defined single-domain LTS state, characterized by six-fold magnetic satellites, coexisting with the TC phase. Together with the magnetic resonance spectroscopy results, which resolve surface and bulk dynamical modes, these findings demonstrate that the Cu$_2$OSeO$_3$/Bi$_2$Se$_3$ interface modifies the near-surface magnetic free-energy landscape. (see Sec.~\ref {REXS} of the Supplemental Material~\cite{supplemental_material}).

%%%%%%%%%

\section{Discussion and Conclusions} \label{sec:conclusions}

In summary, broadband ferromagnetic resonance measurements of the Cu$_2$OSeO$_3$/Bi$_2$Se$_3$ heterostructure reveal pronounced modifications of the magnetic phase behavior compared with bare Cu$_2$OSeO$_3$. In addition to a distinct resonance mode in the TC phase that is absent in bare Cu$_2$OSeO$_3$, field cycling resolves two closely spaced CCW skyrmion resonance branches separated by approximately 238 MHz. The coexistence of these branches is consistent with the presence of bulk and interfacial skyrmion populations experiencing distinct magnetic environments.

%In our study, broadband ferromagnetic resonance of Cu$_2$OSeO$_3$/Bi$_2$Se$_3$\ shows that the Bi$_2$Se$_3$ interface enhances the interfacial DMI and provides a stabilization mechanism for skyrmions, consistent with skyrmion formation reported at TI–magnet interfaces~\cite{yasuda2016geometric,chen2019evidence}, while also giving rise to a new TC mode that is absent in the resonance spectra of bare Cu$_2$OSeO$_3$. Under field cycling we further resolve two closely spaced CCW branches separated by $\sim$238~MHz, which the additional interfacial DMI and exchange coupling naturally explain as surface- and bulk-related skyrmion modes.

Complementary REXS and SQUID magnetometry measurements demonstrate that the skyrmion phase is significantly enhanced at the Cu$_2$OSeO$_3$/Bi$_2$Se$_3$ interface. Field cycling increases the skyrmion population while suppressing competing conical textures, and the resulting sixfold diffraction pattern observed by REXS indicates the formation of a well-ordered low-temperature skyrmion lattice. Furthermore, the reconstructed magnetic phase diagrams reveal an expanded interfacial skyrmion pocket that extends over a broader field range as compared to the corresponding bulk phase and persists into regions where the bulk favors helical order.

Taken together, these observations indicate that proximity effects at the Cu$_2$OSeO$_3$/Bi$_2$Se$_3$ interface modify the magnetic energy landscape and promote the stabilization of topological spin textures. A plausible origin is an interfacial contribution to the effective DMI and exchange coupling arising from the strong spin–orbit coupling of Bi$_2$Se$_3$ and the broken inversion symmetry at the interface. In contrast to conventional heavy-metal-based systems, the Cu$_2$OSeO$_3$/Bi$_2$Se$_3$ heterostructure combines strong interfacial spin–orbit coupling with an insulating magnetic host possessing exceptionally low magnetic damping. These characteristics establish the topological-insulator/chiral-magnet interfaces as a promising platform for engineering chiral spin interactions and extending the stability of skyrmion phases through interface design.

  %I believe leaving the sections in separate files is more organized, change it if you desire 
\section*{Acknowledgements} \label{sec:acknowledgements}
The authors gratefully acknowledge fruitful discussions with M. Garst, T. Hesjedal, and J. Knolle.
This work was supported by the Germany’s Excellence Strategy-[EXC-2111 390814868], [MCQST 528001743].
C. H. Back acknowledges funding by the German Research Foundation via project BA 2181/20-1.
F.R. acknowledges funding by the Deutsche Forschungsgemeinschaft (DFG; German Research Foundation) under SPP2137 Skyrmionics/RA 3570.
M.K. was funded by the DFG - CRC/SFB 1277 Project No. 314695032.
The REXS experiment was conducted at the BESSY II synchrotron, UE51-PGM-1 beamline (Helmholtz-Zentrum Berlin) as part of the proposal "Photons-252-13824-IN/CDA" and by the Deutsche Forschungsgemeinschaft (DFG, German Research Foundation) –[TRR 360–492547816] and DFG grant – 528001743.
We acknowledge financial support for the VEKMAG project by the German Federal Ministry for Education and Research (BMBF 05K2010, 05K2013, 05K2016, 05K2019, 05K2022) and by HZB.

\section{Supplemental Material}

The Supplemental Material provides additional details on the TC mode in the out-of-plane configuration, the influence of the field-cycling protocol on the TC mode, and the temperature dependence (5–25 K) of the in-plane configuration, including the resolution of distinct excitation modes.

\bibliography{sections/bib}

@article{adams2012long,
  title={Long-wavelength helimagnetic order and skyrmion lattice phase in Cu 2 OSeO 3},
  author={Adams, T and Chacon, A and Wagner, M and Bauer, A and Brandl, G and Pedersen, B and Berger, H and Lemmens, P and Pfleiderer, C},
  journal={Physical review letters},
  volume={108},
  number={23},
  pages={237204},
  year={2012},
  publisher={APS}
}

@article{aqeel2022growth,
  title={Growth and helicity of noncentrosymmetric Cu2OSeO3 crystals},
  author={Aqeel, Aisha and Sahliger, Jan and Li, Guowei and Baas, Jacob and Blake, Graeme R and Palstra, Thomas TM and Back, Christian H},
  journal={physica status solidi (b)},
  volume={259},
  number={5},
  pages={2100152},
  year={2022},
  publisher={Wiley Online Library}
}

@article{aqeel2021microwave,
  title={Microwave spectroscopy of the low-temperature skyrmion state in Cu 2 OSeO 3},
  author={Aqeel, Aisha and Sahliger, Jan and Taniguchi, Takuya and M{\"a}ndl, Stefan and Mettus, Denis and Berger, Helmuth and Bauer, Andreas and Garst, Markus and Pfleiderer, Christian and Back, Christian H},
  journal={Physical Review Letters},
  volume={126},
  number={1},
  pages={017202},
  year={2021},
  publisher={APS}
}

@article{azhar2022screw,
  title={Screw dislocations in chiral magnets},
  author={Azhar, Maria and Kravchuk, Volodymyr P and Garst, Markus},
  journal={Physical Review Letters},
  volume={128},
  number={15},
  pages={157204},
  year={2022},
  publisher={APS}
}

@article{bannenberg2019multiple,
  title={Multiple low-temperature skyrmionic states in a bulk chiral magnet},
  author={Bannenberg, Lars J and Wilhelm, Heribert and Cubitt, Robert and Labh, Ankit and Schmidt, Marcus P and Leli{\`e}vre-Berna, Eddy and Pappas, Catherine and Mostovoy, Maxim and Leonov, Andrey O},
  journal={npj Quantum Materials},
  volume={4},
  number={1},
  pages={11},
  year={2019},
  publisher={Nature Publishing Group UK London}
}

@article{camley2023consequences,
  title={Consequences of the Dzyaloshinskii-Moriya interaction},
  author={Camley, Robert E and Livesey, Karen L},
  journal={Surface Science Reports},
  volume={78},
  number={3},
  pages={100605},
  year={2023},
  publisher={Elsevier}
}

@article{chacon2018observation,
  title={Observation of two independent skyrmion phases in a chiral magnetic material},
  author={Chacon, A and Heinen, L and Halder, M and Bauer, A and Simeth, W and M{\"u}hlbauer, S and Berger, H and Garst, Markus and Rosch, Achim and Pfleiderer, Christian},
  journal={Nature physics},
  volume={14},
  number={9},
  pages={936--941},
  year={2018},
  publisher={Nature Publishing Group UK London}
}

@article{chen2019evidence,
  title={Evidence for magnetic skyrmions at the interface of ferromagnet/topological-insulator heterostructures},
  author={Chen, Junshu and Wang, Linjing and Zhang, Meng and Zhou, Liang and Zhang, Runnan and Jin, Lipeng and Wang, Xuesen and Qin, Hailang and Qiu, Yang and Mei, Jiawei and others},
  journal={Nano letters},
  volume={19},
  number={9},
  pages={6144--6151},
  year={2019},
  publisher={ACS Publications}
}

@article{dzyaloshinsky1958thermodynamic,
  title={A thermodynamic theory of “weak” ferromagnetism of antiferromagnetics},
  author={Dzyaloshinsky, Igor},
  journal={Journal of physics and chemistry of solids},
  volume={4},
  number={4},
  pages={241--255},
  year={1958},
  publisher={Elsevier}
}

@article{eremeev2013magnetic,
  title={Magnetic proximity effect at the three-dimensional topological insulator/magnetic insulator interface},
  author={Eremeev, Sergey V and Men'Shov, VN and Tugushev, VV and Echenique, Pedro M and Chulkov, Eugene V},
  journal={Physical Review B—Condensed Matter and Materials Physics},
  volume={88},
  number={14},
  pages={144430},
  year={2013},
  publisher={APS}
}

@article{fanchiang2018strongly,
  title={Strongly exchange-coupled and surface-state-modulated magnetization dynamics in Bi2Se3/yttrium iron garnet heterostructures},
  author={Fanchiang, YT and Chen, KHM and Tseng, CC and Chen, CC and Cheng, CK and Yang, SR and Wu, CN and Lee, SF and Hong, M and Kwo, J},
  journal={Nature communications},
  volume={9},
  number={1},
  pages={223},
  year={2018},
  publisher={Nature Publishing Group UK London}
}

@article{fedel2025evidence,
  title={Evidence of Long-Range Dzyaloshinskii--Moriya Interaction at Ferrimagnetic Insulator/Nonmagnetic Metal Interfaces},
  author={Fedel, Stefano and Villa, Mario and Damerio, Silvia and Demiroglu, Emre and Deger, Caner and Gazquez, Jaume and Avci, Can O},
  journal={Advanced Functional Materials},
  pages={2418653},
  year={2025},
  publisher={Wiley Online Library}
}

@article{fert2017magnetic,
  title={Magnetic skyrmions: advances in physics and potential applications},
  author={Fert, Albert and Reyren, Nicolas and Cros, Vincent},
  journal={Nature Reviews Materials},
  volume={2},
  number={7},
  pages={1--15},
  year={2017},
  publisher={Nature Publishing Group}
}

@article{gobel2021beyond,
  title={Beyond skyrmions: Review and perspectives of alternative magnetic quasiparticles},
  author={G{\"o}bel, B{\"o}rge and Mertig, Ingrid and Tretiakov, Oleg A},
  journal={Physics Reports},
  volume={895},
  pages={1--28},
  year={2021},
  publisher={Elsevier}
}

@article{halder2018thermodynamic,
  title={Thermodynamic evidence of a second skyrmion lattice phase and tilted conical phase in Cu 2 OSeO 3},
  author={Halder, M and Chacon, A and Bauer, A and Simeth, W and M{\"u}hlbauer, S and Berger, H and Heinen, L and Garst, M and Rosch, A and Pfleiderer, C},
  journal={Physical review B},
  volume={98},
  number={14},
  pages={144429},
  year={2018},
  publisher={APS}
}

@article{he2017tailoring,
  title={Tailoring exchange couplings in magnetic topological-insulator/antiferromagnet heterostructures},
  author={He, Qing Lin and Kou, Xufeng and Grutter, Alexander J and Yin, Gen and Pan, Lei and Che, Xiaoyu and Liu, Yuxiang and Nie, Tianxiao and Zhang, Bin and Disseler, Steven M and others},
  journal={Nature materials},
  volume={16},
  number={1},
  pages={94--100},
  year={2017},
  publisher={Nature Publishing Group UK London}
}

@article{huang2018situ,
  title={In situ electric field skyrmion creation in magnetoelectric Cu2OSeO3},
  author={Huang, Ping and Cantoni, Marco and Kruchkov, Alex and Rajeswari, Jayaraman and Magrez, Arnaud and Carbone, Fabrizio and R{\o}nnow, Henrik M},
  journal={Nano letters},
  volume={18},
  number={8},
  pages={5167--5171},
  year={2018},
  publisher={ACS Publications}
}

@article{katmis2016high,
  title={A high-temperature ferromagnetic topological insulating phase by proximity coupling},
  author={Katmis, Ferhat and Lauter, Valeria and Nogueira, Flavio S and Assaf, Badih A and Jamer, Michelle E and Wei, Peng and Satpati, Biswarup and Freeland, John W and Eremin, Ilya and Heiman, Don and others},
  journal={Nature},
  volume={533},
  number={7604},
  pages={513--516},
  year={2016},
  publisher={Nature Publishing Group}
}

@article{koraltan20262026,
  title={The 2026 Skyrmionics Roadmap},
  author={Koraltan, Sabri and Abert, Claas and Albrecht, Manfred and Azhar, Maria and Back, Christian and B{\'e}a, H{\'e}l{\`e}ne and Birch, Max T and Bl{\"u}gel, Stefan and Boulle, Olivier and B{\"u}ttner, Felix and others},
  journal={arXiv preprint arXiv:2601.16575},
  year={2026}
}

@article{lee2021tunable,
  title={Tunable gigahertz dynamics of low-temperature skyrmion lattice in a chiral magnet},
  author={Lee, Oscar and Sahliger, Jan and Aqeel, Aisha and Khan, Safe and Seki, Shinichiro and Kurebayashi, Hidekazu and Back, Christian H},
  journal={Journal of Physics: Condensed Matter},
  volume={34},
  number={9},
  pages={095801},
  year={2021},
  publisher={IOP Publishing}
}

@article{lee2024task,
  title={Task-adaptive physical reservoir computing},
  author={Lee, Oscar and Wei, Tianyi and Stenning, Kilian D and Gartside, Jack C and Prestwood, Dan and Seki, Shinichiro and Aqeel, Aisha and Karube, Kosuke and Kanazawa, Naoya and Taguchi, Yasujiro and others},
  journal={Nature materials},
  volume={23},
  number={1},
  pages={79--87},
  year={2024},
  publisher={Nature Publishing Group UK London}
}

@article{li2020topological,
  title={Topological Hall effect in a topological insulator interfaced with a magnetic insulator},
  author={Li, Peng and Ding, Jinjun and Zhang, Steven S-L and Kally, James and Pillsbury, Timothy and Heinonen, Olle G and Rimal, Gaurab and Bi, Chong and DeMann, August and Field, Stuart B and others},
  journal={Nano letters},
  volume={21},
  number={1},
  pages={84--90},
  year={2020},
  publisher={ACS Publications}
}

@article{liu2020changes,
  title={Changes of Magnetism in a Magnetic Insulator due to Proximity to a Topological Insulator},
  author={Liu, Tao and Kally, James and Pillsbury, Timothy and Liu, Chuanpu and Chang, Houchen and Ding, Jinjun and Cheng, Yang and Hilse, Maria and Engel-Herbert, Roman and Richardella, Anthony and others},
  journal={Physical review letters},
  volume={125},
  number={1},
  pages={017204},
  year={2020},
  publisher={APS}
}

@article{luthi2023hybrid,
  title={Hybrid magnetization dynamics in Cu2OSeO3/NiFe heterostructures},
  author={L{\"u}thi, Carolina and Flacke, Luis and Aqeel, Aisha and Kamra, Akashdeep and Gross, Rudolf and Back, Christian and Weiler, Mathias},
  journal={Applied Physics Letters},
  volume={122},
  number={1},
  year={2023},
  publisher={AIP Publishing}
}

@article{lv2018unidirectional,
  title={Unidirectional spin-Hall and Rashba- Edelstein magnetoresistance in topological insulator-ferromagnet layer heterostructures},
  author={Lv, Yang and Kally, James and Zhang, Delin and Lee, Joon Sue and Jamali, Mahdi and Samarth, Nitin and Wang, Jian-Ping},
  journal={Nature communications},
  volume={9},
  number={1},
  pages={111},
  year={2018},
  publisher={Nature Publishing Group UK London}
}

@article{lv2022large,
  title={Large unidirectional spin Hall and Rashba- Edelstein magnetoresistance in topological insulator/magnetic insulator heterostructures},
  author={Lv, Yang and Kally, James and Liu, Tao and Quarterman, Patrick and Pillsbury, Timothy and Kirby, Brian J and Grutter, Alexander J and Sahu, Protyush and Borchers, Julie A and Wu, Mingzhong and others},
  journal={Applied Physics Reviews},
  volume={9},
  number={1},
  year={2022},
  publisher={AIP Publishing}
}

@article{maier2018note,
  title={Note: Derivative divide, a method for the analysis of broadband ferromagnetic resonance in the frequency domain},
  author={Maier-Flaig, Hannes and Goennenwein, Sebastian TB and Ohshima, Ryo and Shiraishi, Masashi and Gross, Rudolf and Huebl, Hans and Weiler, Mathias},
  journal={Review of Scientific Instruments},
  volume={89},
  number={7},
  year={2018},
  publisher={AIP Publishing}
}

@article{marchiori2024imaging,
  title={Imaging magnetic spiral phases, skyrmion clusters, and skyrmion displacements at the surface of bulk Cu2OSeO3},
  author={Marchiori, Estefani and Romagnoli, Giulio and Schneider, Lukas and Gross, Boris and Sahafi, Pardis and Jordan, Andrew and Budakian, Raffi and Baral, Priya R and Magrez, Arnaud and White, Jonathan S and others},
  journal={Communications Materials},
  volume={5},
  number={1},
  pages={202},
  year={2024},
  publisher={Nature Publishing Group UK London}
}

@article{mehboodi2024observation,
  title={Observation of distorted tilted conical phase at the surface of a bulk chiral magnet with resonant elastic x-ray scattering},
  author={Mehboodi, S and Ukleev, V and Luo, C and Abrudan, R and Radu, F and Back, CH and Aqeel, A},
  journal={arXiv preprint arXiv:2412.15882},
  year={2024}
}

@article{mochizuki2012spin,
  title={Spin-wave modes and their intense excitation effects in skyrmion crystals},
  author={Mochizuki, Masahito},
  journal={Physical review letters},
  volume={108},
  number={1},
  pages={017601},
  year={2012},
  publisher={APS}
}

@article{moriya1960anisotropic,
  title={Anisotropic superexchange interaction and weak ferromagnetism},
  author={Moriya, T{\^o}ru},
  journal={Physical review},
  volume={120},
  number={1},
  pages={91},
  year={1960},
  publisher={APS}
}

@article{muhlbauer2009skyrmion,
  title={Skyrmion lattice in a chiral magnet},
  author={M\"uhlbauer, Sebastian and Binz, Benedikt and Jonietz, Florian and Pfleiderer, Christian and Rosch, Achim and Neubauer, Anja and Georgii, Robert and B\"oni, Peter},
  journal={Science},
  volume={323},
  number={5916},
  pages={915--919},
  year={2009},
  publisher={American Association for the Advancement of Science}
}

@article{nagaosa2013topological,
  title={Topological properties and dynamics of magnetic skyrmions},
  author={Nagaosa, Naoto and Tokura, Yoshinori},
  journal={Nature nanotechnology},
  volume={8},
  number={12},
  pages={899--911},
  year={2013},
  publisher={Nature Publishing Group UK London}
}

@article{navabi2019control,
  title={Control of spin-wave damping in YIG using spin currents from topological insulators},
  author={Navabi, Aryan and Liu, Yuxiang and Upadhyaya, Pramey and Murata, Koichi and Ebrahimi, Farbod and Yu, Guoqiang and Ma, Bo and Rao, Yiheng and Yazdani, Mohsen and Montazeri, Mohammad and others},
  journal={Physical review applied},
  volume={11},
  number={3},
  pages={034046},
  year={2019},
  publisher={APS}
}

@article{okamura2013microwave,
  title={Microwave magnetoelectric effect via skyrmion resonance modes in a helimagnetic multiferroic},
  author={Okamura, Y and Kagawa, F and Mochizuki, M and Kubota, M and Seki, S and Ishiwata, S and Kawasaki, M and Onose, Y and Tokura, Y},
  journal={Nature Communications},
  volume={4},
  number={1},
  pages={2391},
  year={2013},
  publisher={Nature Publishing Group UK London}
}

@article{onose2012observation,
  title={Observation of magnetic excitations of skyrmion crystal in a helimagnetic insulator Cu 2 OSeO 3},
  author={Onose, Y and Okamura, Y and Seki, S and Ishiwata, S and Tokura, Y},
  journal={Physical review letters},
  volume={109},
  number={3},
  pages={037603},
  year={2012},
  publisher={APS}
}

@article{qian2018new,
  title={New magnetic phase of the chiral skyrmion material Cu2OSeO3},
  author={Qian, Fengjiao and Bannenberg, Lars J and Wilhelm, Heribert and Chaboussant, Gr{\'e}gory and Debeer-Schmitt, Lisa M and Schmidt, Marcus P and Aqeel, Aisha and Palstra, Thomas TM and Br{\"u}ck, Ekkes and Lefering, Anton JE and others},
  journal={Science Advances},
  volume={4},
  number={9},
  pages={eaat7323},
  year={2018},
  publisher={American Association for the Advancement of Science}
}

@article{ran2021creation,
  title={Creation of a chiral bobber lattice in helimagnet-multilayer heterostructures},
  author={Ran, Kejing and Liu, Yizhou and Guang, Yao and Burn, David M and van der Laan, Gerrit and Hesjedal, Thorsten and Du, Haifeng and Yu, Guoqiang and Zhang, Shilei},
  journal={Physical Review Letters},
  volume={126},
  number={1},
  pages={017204},
  year={2021},
  publisher={APS}
}

@article{schwarze2015universal,
  title={Universal helimagnon and skyrmion excitations in metallic, semiconducting and insulating chiral magnets},
  author={Schwarze, T and Waizner, Johannes and Garst, Markus and Bauer, Andreas and Stasinopoulos, Ioannis and Berger, Helmuth and Pfleiderer, Christian and Grundler, Dirk},
  journal={Nature materials},
  volume={14},
  number={5},
  pages={478--483},
  year={2015},
  publisher={Nature Publishing Group UK London}
}

@article{seki2012observation,
  title={Observation of skyrmions in a multiferroic material},
  author={Seki, Shinichiro and Yu, XZ and Ishiwata, S and Tokura, Yoshinori},
  journal={Science},
  volume={336},
  number={6078},
  pages={198--201},
  year={2012},
  publisher={American Association for the Advancement of Science}
}

@article{singh2024anisotropic,
  title={Anisotropic planar Hall effects in Bi 2 Se 3/EuS interfaces: Deciphering the role of proximity-induced spin canting and topological spin texture},
  author={Singh, Juhi and Raman, Karthik V and Mohanta, Narayan},
  journal={Physical Review B},
  volume={110},
  number={12},
  pages={125133},
  year={2024},
  publisher={APS}
}

@article{stasinopoulos2017low,
  title={Low spin wave damping in the insulating chiral magnet Cu2OSeO3},
  author={Stasinopoulos, I and Weichselbaumer, S and Bauer, A and Waizner, J and Berger, H and Maendl, S and Garst, M and Pfleiderer, C and Grundler, D},
  journal={Applied Physics Letters},
  volume={111},
  number={3},
  year={2017},
  publisher={AIP Publishing}
}

@article{versteeg2016optically,
  title={Optically probed symmetry breaking in the chiral magnet Cu 2 OSeO 3},
  author={Versteeg, RB and Vergara, I and Sch{\"a}fer, SD and Bischoff, D and Aqeel, Aisha and Palstra, TTM and Gr{\"u}ninger, M and Van Loosdrecht, PHM},
  journal={Physical Review B},
  volume={94},
  number={9},
  pages={094409},
  year={2016},
  publisher={APS}
}

@article{wang2016surface,
  title={Surface-state-dominated spin-charge current conversion in topological-insulator--ferromagnetic-insulator heterostructures},
  author={Wang, Hailong and Kally, James and Lee, Joon Sue and Liu, Tao and Chang, Houchen and Hickey, Danielle Reifsnyder and Mkhoyan, K Andre and Wu, Mingzhong and Richardella, Anthony and Samarth, Nitin},
  journal={Physical review letters},
  volume={117},
  number={7},
  pages={076601},
  year={2016},
  publisher={APS}
}

@article{wang2022observation,
  title={Observation of nonlinear planar Hall effect in magnetic-insulator--topological-insulator heterostructures},
  author={Wang, Yang and Mambakkam, Sivakumar V and Huang, Yue-Xin and Wang, Yong and Ji, Yi and Xiao, Cong and Yang, Shengyuan A and Law, Stephanie A and Xiao, John Q},
  journal={Physical Review B},
  volume={106},
  number={15},
  pages={155408},
  year={2022},
  publisher={APS}
}

@article{white2014electric,
  title={Electric-field-induced skyrmion distortion and giant lattice rotation in the magnetoelectric insulator Cu 2 OSeO 3},
  author={White, JS and Pr{\v{s}}a, K and Huang, P and Omrani, AA and {\v{Z}}ivkovi{\'c}, I and Bartkowiak, M and Berger, H and Magrez, A and Gavilano, JL and Nagy, G and others},
  journal={Physical review letters},
  volume={113},
  number={10},
  pages={107203},
  year={2014},
  publisher={APS}
}

@article{yasuda2016geometric,
  title={Geometric Hall effects in topological insulator heterostructures},
  author={Yasuda, K and Wakatsuki, R and Morimoto, T and Yoshimi, R and Tsukazaki, A and Takahashi, KS and Ezawa, M and Kawasaki, M and Nagaosa, N and Tokura, Y},
  journal={Nature Physics},
  volume={12},
  number={6},
  pages={555--559},
  year={2016},
  publisher={Nature Publishing Group UK London}
}

@article{yu2010real,
  title={Real-space observation of a two-dimensional skyrmion crystal},
  author={Yu, XZ and Onose, Yoshinori and Kanazawa, Naoya and Park, Joung Hwan and Han, JH and Matsui, Yoshio and Nagaosa, Naoto and Tokura, Yoshinori},
  journal={Nature},
  volume={465},
  number={7300},
  pages={901--904},
  year={2010},
  publisher={Nature Publishing Group UK London}
}

@article{zhang2015magnetic,
  title={Magnetic skyrmion logic gates: conversion, duplication and merging of skyrmions},
  author={Zhang, Xichao and Ezawa, Motohiko and Zhou, Yan},
  journal={Scientific reports},
  volume={5},
  number={1},
  pages={1--8},
  year={2015},
  publisher={Nature Publishing Group}
}

@article{zhu2018proximity,
  title={Proximity-induced magnetism and an anomalous Hall effect in Bi 2 Se 3/LaCoO 3: A topological insulator/ferromagnetic insulator thin film heterostructure},
  author={Zhu, Shanna and Meng, Dechao and Liang, Genhao and Shi, Gang and Zhao, Peng and Cheng, Peng and Li, Yongqing and Zhai, Xiaofang and Lu, Yalin and Chen, Lan and others},
  journal={Nanoscale},
  volume={10},
  number={21},
  pages={10041--10049},
  year={2018},
  publisher={Royal Society of Chemistry}
}

@misc{supplemental_material, note = {See Supplemental Material at [URL will be inserted by publisher] for experimental details and supporting measurements} }

\clearpage
\onecolumngrid
\newpage
\setcounter{page}{1}
\pagenumbering{arabic}
\begin{center}
{\large\bfseries Supplemental Material for}\\[0.5em]
{\large\bfseries
Proximity-Induced Skyrmion Stabilization at the
Cu$_2$OSeO$_3$/Bi$_2$Se$_3$ Interface
}\\[1em]

S.~Mehboodi$^{1,2,3}$,
V.~Ukleev$^{4}$,
C.~Luo$^{4}$,
R.~Abrudan$^{4}$,
J.~Xiao$^{4}$,
R.~Golnak$^{4}$,
F.~Radu$^{4}$,
M.~Kronseder$^{5}$,
C.~H.~Back$^{1,2,3}$,
and A.~Aqeel$^{1,2,6}$

\vspace{0.8em}

{\small
$^{1}$School of Natural Sciences, Technical University of Munich,
85748 Garching, Germany\\
$^{2}$Munich Center for Quantum Science and Technology (MCQST),
Munich, Germany\\
$^{3}$Center for Quantum Engineering (ZQE),
Technical University of Munich, 85748 Garching, Germany\\
$^{4}$Helmholtz-Zentrum Berlin für Materialien und Energie,
Berlin, Germany\\
$^{5}$Institute for Experimental and Applied Physics,
University of Regensburg, 93040 Regensburg, Germany\\
$^{6}$Institute of Physics, University of Augsburg,
86135 Augsburg, Germany
}
\end{center}

\vspace{1em}

\setcounter{section}{0}
\setcounter{figure}{0}
\setcounter{table}{0}
\setcounter{equation}{0}

\renewcommand{\thesection}{S\arabic{section}}
\renewcommand{\thesubsection}{S\arabic{section}.\arabic{subsection}}
\renewcommand{\thefigure}{S\arabic{figure}}
\renewcommand{\thetable}{S\arabic{table}}
\renewcommand{\theequation}{S\arabic{equation}}

\renewcommand{\thesection}{S\arabic{section}}
\setcounter{section}{0}
% \section*{Supplemental Material}
\section{Resonance modes detected in out-of-plane configuration} \label{sec:appendix}

\begin{figure*}[ht]
    \centering
    \includegraphics[]{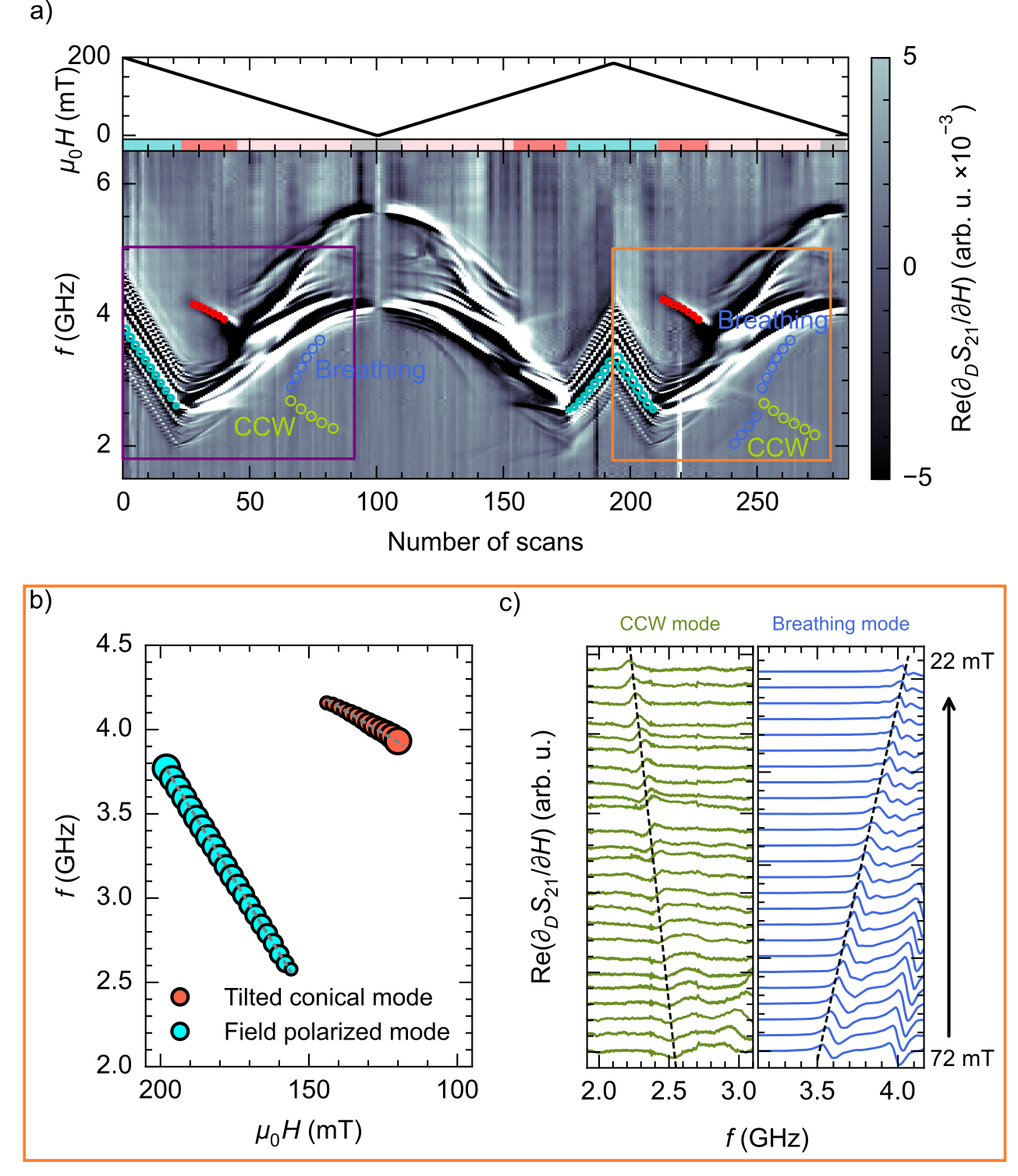}
    \caption{(a) Experimental spectra and fits of different magnetic modes at 5~K. The top panel shows the magnetic field sweep protocol. The green, cyan, red, and blue circles highlight the CCW, FP, tilted conical (TC), and breathing modes, respectively. (b) represents the extracted resonance frequency of the FP (cyan) and the TC (red) modes for the right side of the spectrum enclosed by an orange rectangle (field sweeping from FP to zero, the purple rectangle marks the equivalent region of the first down‑sweep). (c) shows the line scans as a function of frequency, highlighting CCW (green) and breathing (blue) modes of skyrmion for the right sides of experimental spectra. The black dashed lines highlight the field evolution of the resonance frequency.}
    \label{app_spec}
\end{figure*}

Figure~\ref{app_spec}(a) presents the FMR spectra measured on the Cu$_2$OSeO$_3$/Bi$_2$Se$_3$ heterostructure under a continuous magnetic field sweep. The field was swept in a sequence starting from FP state at 200~mT, to zero, back to FP, and finally brought back to zero. This protocol enables tracking the evolution of magnetic modes across successive field sweeps without applying a separate cycling routine.
The top panel displays the magnetic field as a function of scan number. Distinct branches corresponding to different magnetic modes are visible, where the green markers denote the CCW skyrmion mode, while blue highlights the breathing mode. The FP mode (Kittel mode) is marked in cyan, and an additional mode emerging near the FP-to-skyrmion transition is indicated in red.

Importantly, a new well-ordered mode appears just after passing the critical field $\mu_0H_\text{c2} \approx 150$~mT. This mode emerges with a different slope (same slope sign) to the Kittel mode (see Fig.~\ref{app_spec}(b)) and evolves continuously until it intersects the $\pm$Q modes of the conical state. It spans a frequency range from approximately 3.7~GHz to 4.3~GHz and occurs within a magnetic field window of about 150~mT to 115~mT, the vicinity where the TC spiral mode is expected (region with mixed phases). The appearance of this mode likely originates from an enhancement of the interfacial DMI due to the proximity effect of the topological insulator, which stabilizes a more robust magnetic texture distinctly observed in the FMR spectra.
Panel (c) presents frequency-dependent line scans of the FMR spectra obtained from the right-hand orange rectangular region, with the magnetic field swept from the FP state to zero. 
The CCW and breathing modes are visible between 72 and 22~mT, and the black dashed lines trace their resonance peaks, which shift to lower and higher frequencies, respectively, as the field decreases.

\subsection{Effect of field cycling protocol}

\begin{figure*}[ht]
    \centering
    \includegraphics[]{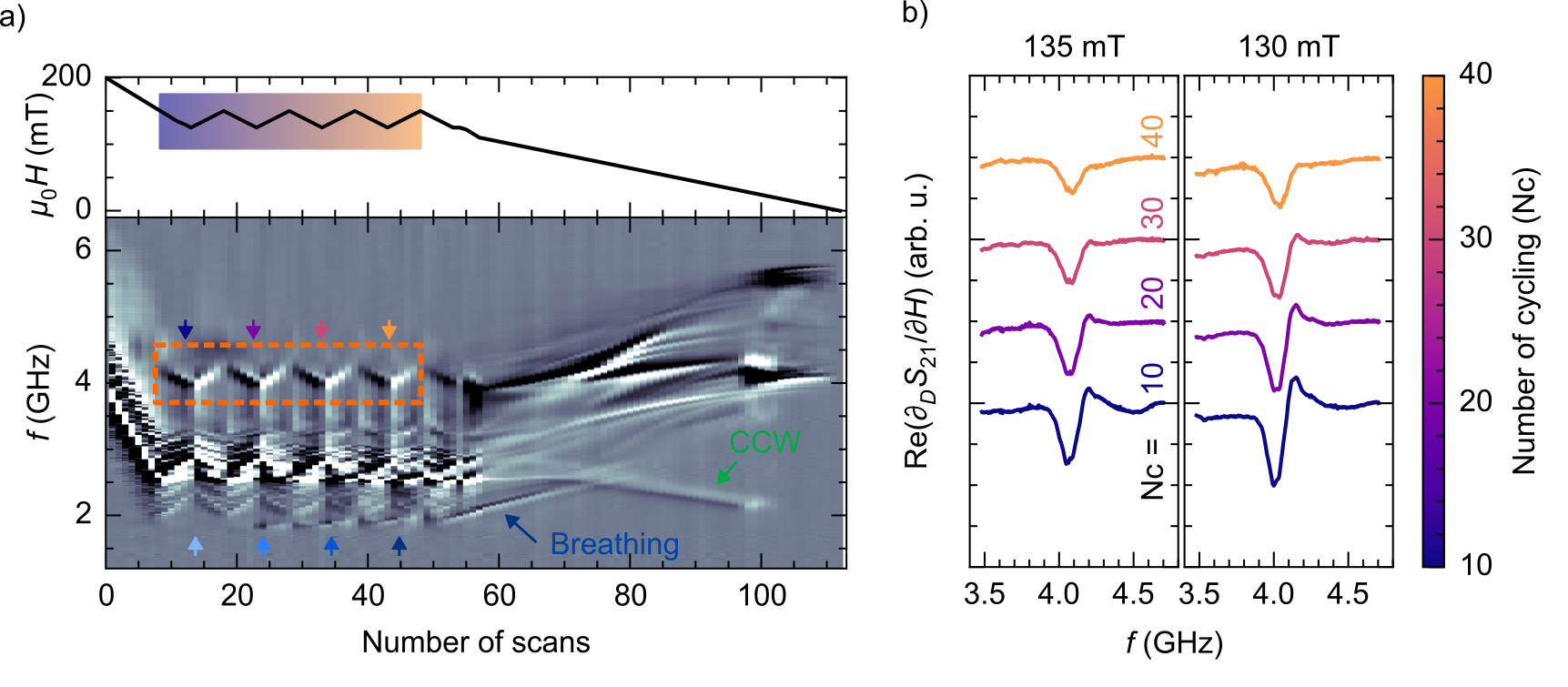}
    \caption{(a) Experimental spectra obtained under cycling at 5~K with magnetic field \(\mu_0H\) perpendicular to the sample's surface. The orange rectangle highlights the region with the well-ordered TC mode with 40 cycles (every 10th cycle is recorded here, further highlighted by arrows on top). The gradient blue arrows on the bottom show the evolution and change of skyrmion mode amplitude for each cycling segment (the lighter the color, the lower the amplitude). (b) Field cuts at two different magnetic field strengths, showing the influence of the cycling protocol on the well-ordered TC mode. Each trace is color-coded as a function of the number of field cycles.}
    \label{cycle_unk}
\end{figure*}

To further investigate the effect of magnetic-field cycling on the mode dynamics, we analyzed how repeated cycling modifies the spectral features of the TC mode. 
The evolution of the mode amplitude provides insight into the stability and redistribution of magnetic textures under periodic driving. 
Figure~\ref{cycle_unk}(a) presents the experimental FMR spectra, starting from a magnetic field of 200~mT (field-polarized state) and continuing beyond $\mu_0H_{c2}$, followed by 40 cycles between $\mu_0H_{\text{low}} = 125$~mT and $\mu_0H_{\text{high}} = 150$~mT (scan 13 to 53). From scan 54, by decreasing the magnetic field to zero, the spectra exhibit conical modes as well as CCW and breathing modes in the lower frequency range. The gradual change from light to dark blue arrows indicates the emergence and increasing amplitude of the breathing mode.
The orange dashed rectangle marks the field-cycled region, highlighting the presence of the TC mode and its evolution during the cycling protocol. 
Note that only one spectrum was recorded after every ten cycles.

As shown in figure~\ref {cycle_unk}(b), frequency-dependent line scans were extracted at two magnetic fields (135~mT and 130~mT, shown in separate panels) after every ten-cycle increment in the field-cycling protocol, corresponding to measurements after 10, 20, 30, and 40 total cycles. 
Each trace is color-coded according to the number of cycles ($N_\mathrm{c}$). 
In both panels, the amplitude of the TC mode gradually decreases with increasing $N_\mathrm{c}$, indicating its sensitivity to repeated cycling. 
This suppression in amplitude, together with the enhanced skyrmion-mode signal, suggests that the TC mode shares a similar dynamic response to the cycling field as the conical mode. 
The trend may also reflect interfacial effects induced by the Bi$_2$Se$_3$ layer, which influence mode stability and skyrmion population.

\begin{figure*}
    \centering
    \includegraphics[]{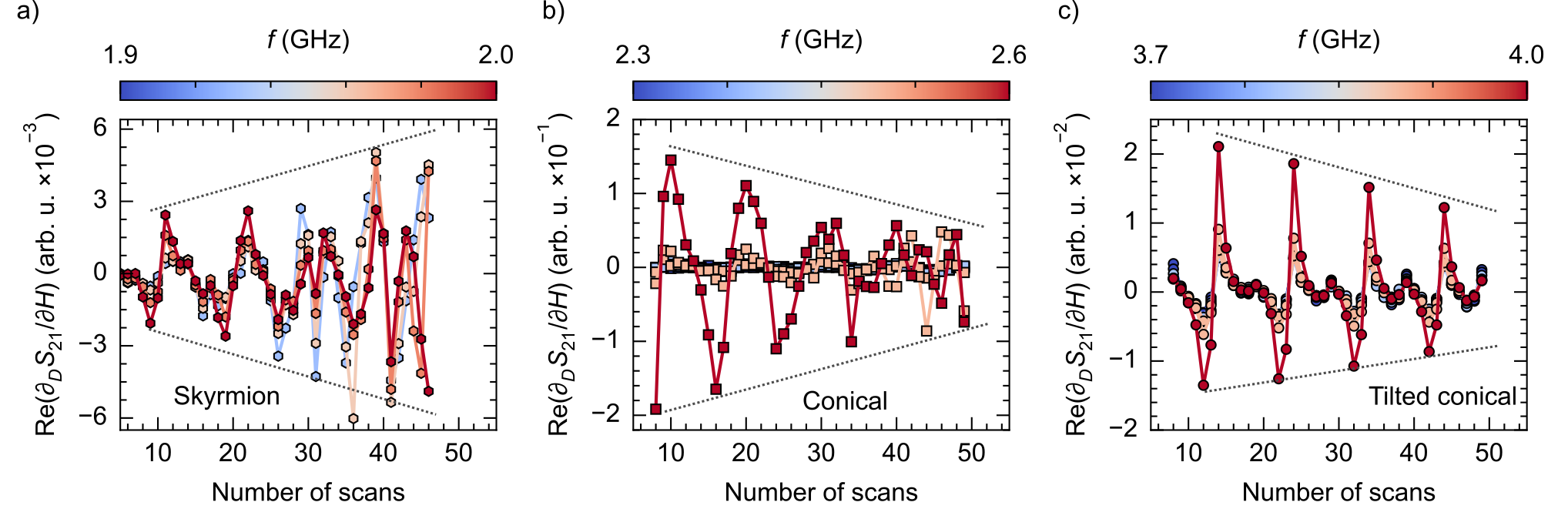}
    \caption{Frequency line cuts as a function of scan number for selected frequency ranges corresponding to different spectral regions. Each panel highlights the evolution of signal amplitude under the cycling protocol for distinct magnetic phases: skyrmion (left), conical (center), and TC (right) modes. The progressive change in amplitude illustrates the phase-specific response to repeated cycling, with dashed gray lines serving as guides to the eye for the envelope of the modulation.}
    \label{freq_cut}
\end{figure*}

Another procedure to visualize how the magnetic modes evolve under repeated field cycling,
figure~\ref{freq_cut} presents frequency line cuts of the FMR spectra in figure~\ref{cycle_unk}(a), plotted as a function of scan number for selected frequencies associated with three distinct magnetic phases: skyrmion (left), conical (middle), and TC mode (right). Each panel represents a fixed-frequency slice extracted from the field-cycled spectra shown in figure~\ref{cycle_unk}, enabling a direct comparison of how the amplitude of each mode evolves under repeated cycling.
For the conical mode, a clear modulation pattern is observed, with amplitude oscillations diminishing throughout the scans. This decay is highlighted by the narrowing envelope (dashed gray lines). The skyrmion mode (middle panel) displays a reverse pattern, though with increasing amplitude modulation across scans, consistent with the formation and stabilization of skyrmions under cycling. In contrast, the TC mode (right) shows a rapid initial decrease in amplitude, with multiple frequency traces collapsing into a narrower envelope over the course of the scans.
Together, these trends demonstrate a phase-specific response to the magnetic field cycling. The pronounced envelope decay in both the conical and the new mode (TC) suggests a transition away from these configurations as skyrmion formation becomes dominant. These dynamics provide further evidence that this new mode (TC) acts as a transitional feature sensitive to the evolving spin texture and likely influenced by interfacial coupling effects in the heterostructure.

\section{Resonance modes detected in in-plane configuration}
\subsection{Temperature dependence of resonance modes}\label{sec:tdependency}
\begin{figure}[ht]
    \centering
    \includegraphics[]{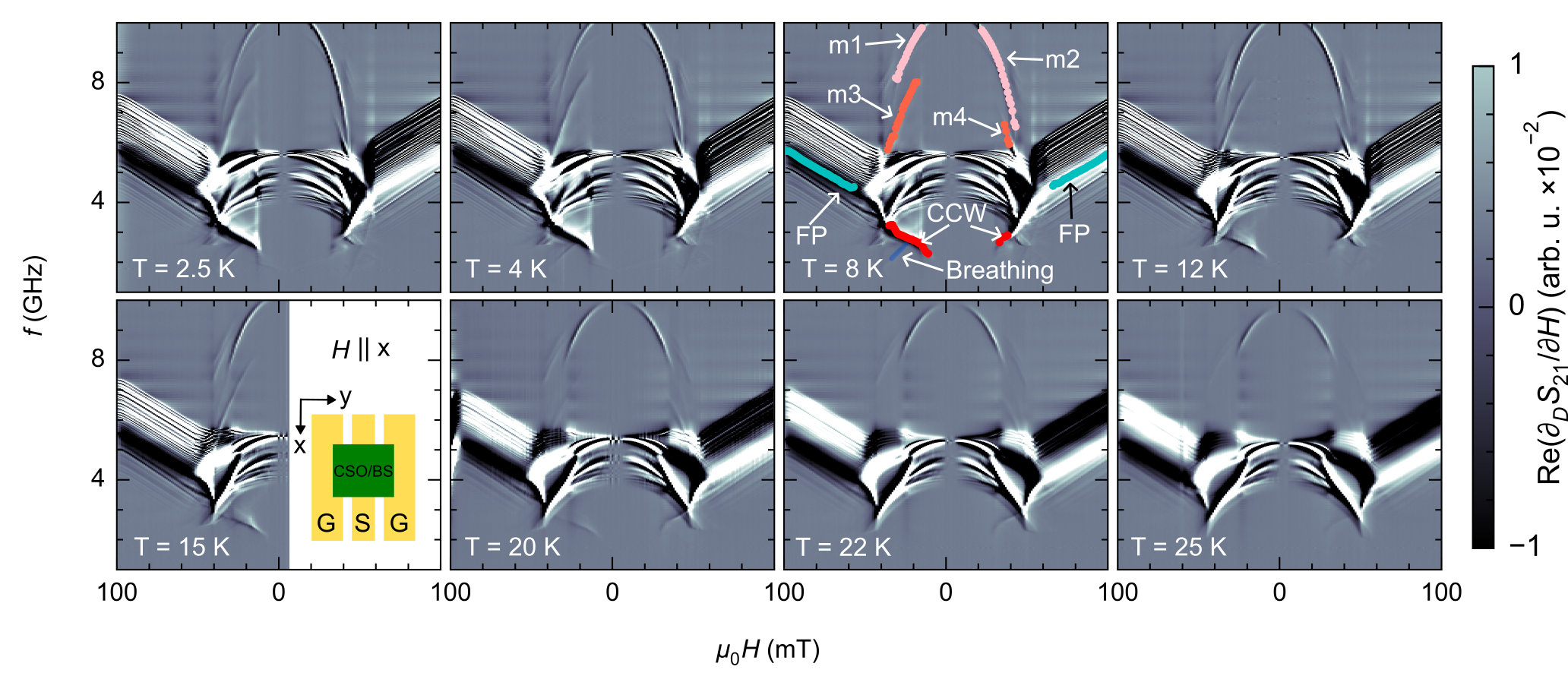}
    \caption{Microwave transmission spectra of the  Cu$_2$OSeO$_3$/Bi$_2$Se$_3$ sample, showing collective spin excitation modes at eight different temperatures. The magnetic field \(\mu_0H\) is swept from the FP state down to zero and then back to the FP state. The spectra reveal the temperature evolution of magnetic resonance features, particularly the LTS mode. 
    The third panel measured at 8K, highlights fitted
    % The 8~K panel includes extracted 
    resonance frequencies corresponding to distinct spin excitation modes, including FP mode, CCW mode, and additional modes (m1-m4). The inset shows schematics of the measurement configuration, with the sample placed on a CPW and the magnetic field applied in-plane, \(H \parallel {\bf x}\).}
    \label{ip-specs}
\end{figure}
Figure~\ref{ip-specs} presents microwave transmission spectra of the  Cu$_2$OSeO$_3$/Bi$_2$Se$_3$ sample, measured using VNA-FMR at eight different temperatures. The inset illustrates the experimental configuration, with the sample mounted on a CPW and the magnetic field applied in-plane along the x-direction. The magnetic field \(\mu_0H\) was swept from the FP state down to zero and then increased back to the FP state, enabling the tracking of magnetic resonance features during both decreasing and increasing field sweeps.

The spectra reveal a strong temperature dependence of the magnetic excitation modes, with noticeable variations in mode positions and intensities. At lower temperatures, the modes become more pronounced, suggesting enhanced stability of the corresponding magnetic phases, particularly the counterclockwise (CCW) mode, as well as the standing spin waves (SSWs) modes observed alongside the Kittel mode. In the 8~K panel, several distinct resonance modes are fitted and labeled, including the FP mode, the CCW mode, and a set of higher-frequency modes m\(_1\) to m\(_4\)). These temperature-dependent measurements were used to reconstruct the phase diagram shown in the main manuscript.

Interestingly, the CCW mode observed during the increasing-field sweep (right side of the spectra) remains visible up to 25~K, indicating a broader temperature stability range compared to the CCW mode in the decreasing-field sweep (left side), which vanishes at lower temperatures (22~K). However, the CCW mode during the decreasing-field sweep appears over a wider field range than its counterpart during the increasing sweep. Furthermore, the m\(_1\) and m\(_2\) modes closely follow the field ranges of the CCW modes in the left and right sides of the spectra, respectively, suggesting a possible correlation between these excitations.

\begin{figure}
    \centering
    \includegraphics[]{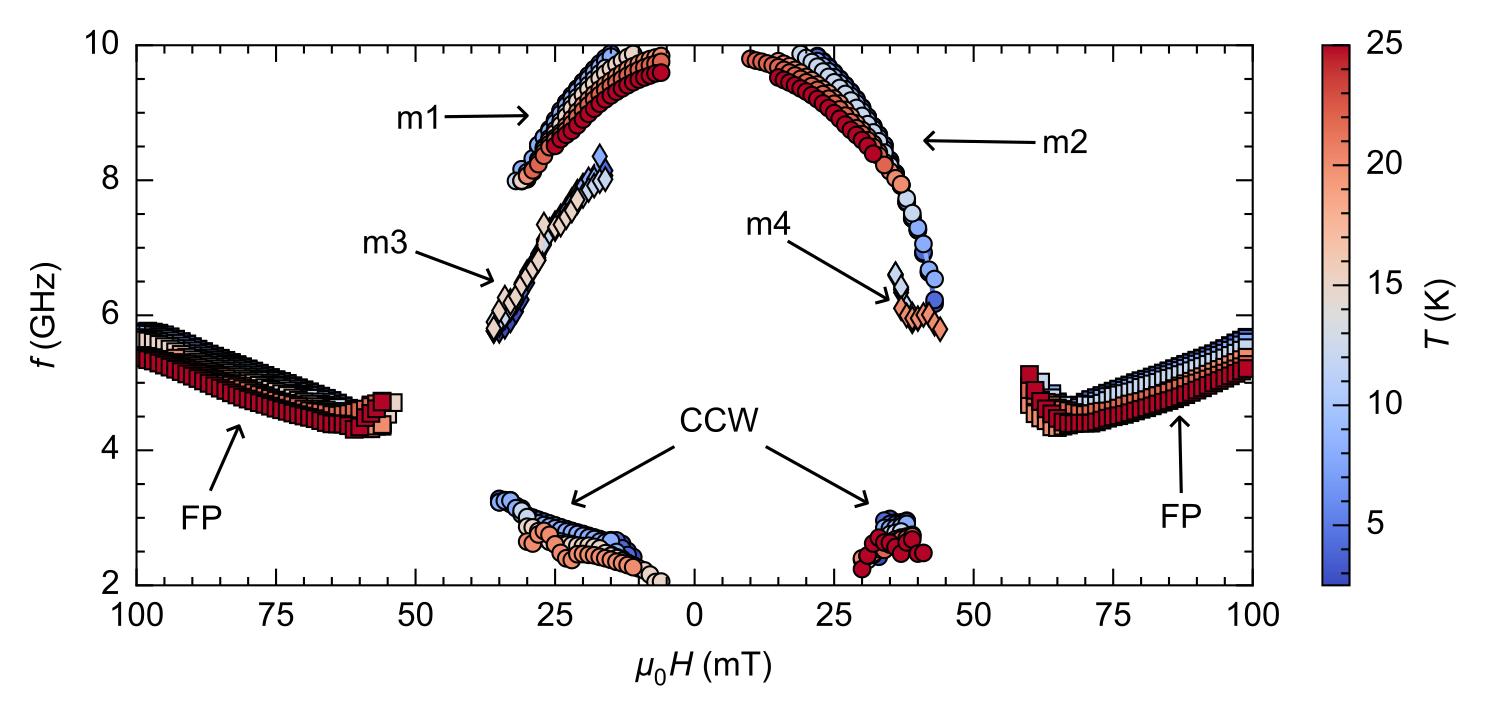}
    \caption{Temperature dependence of the spin excitation spectra measured in the in-plane magnetic field configuration in Fig.~\ref{ip-specs}.}
    \label{fig:Tdependent}
\end{figure}

\begin{figure}
    \centering
    \includegraphics[]{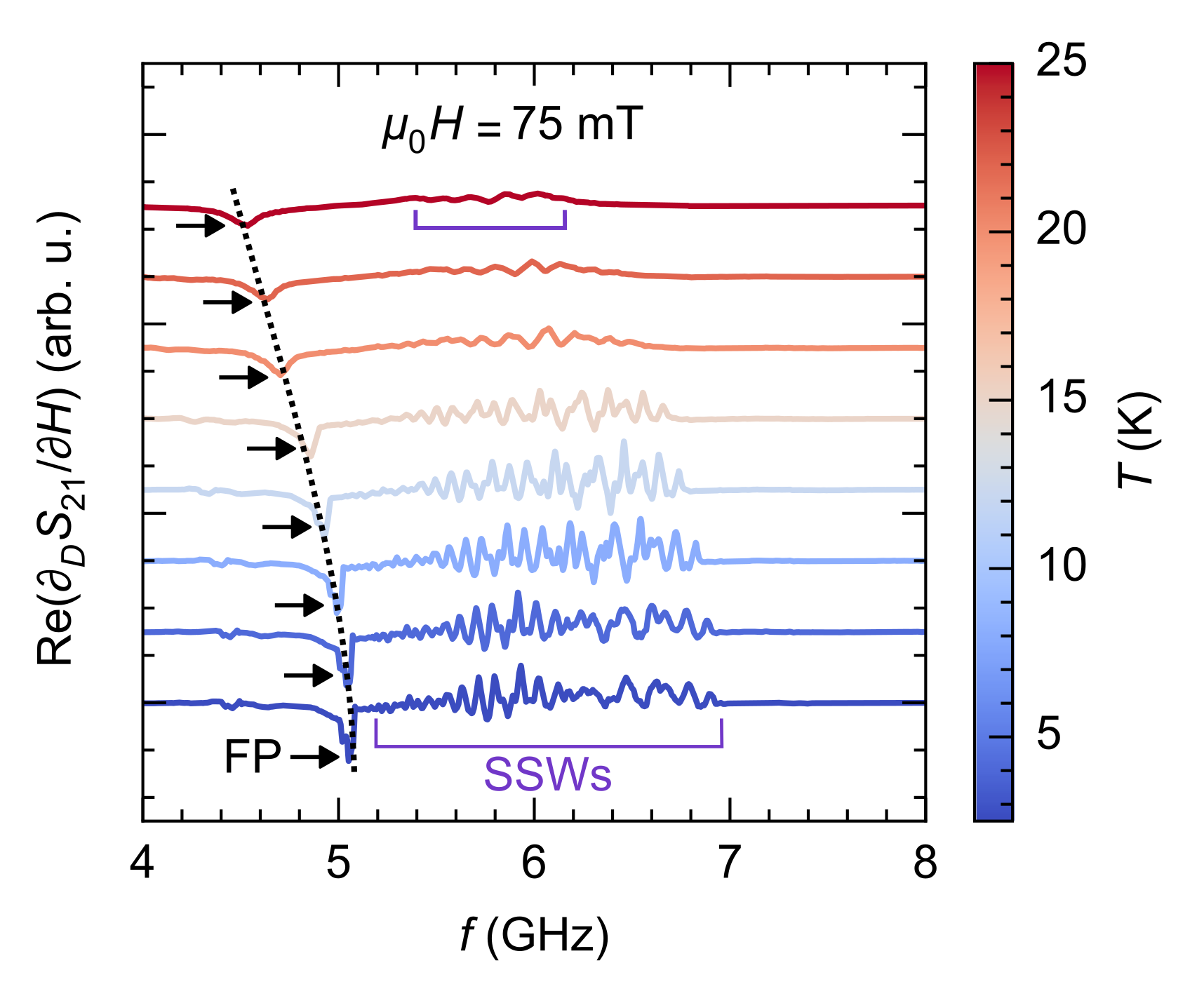}
    \caption{Microwave spectra measured at a fixed magnetic field of \(\mu_0H\)=~75~mT, showing the temperature evolution of spin excitations in the field-polarized phase of the  Cu$_2$OSeO$_3$/Bi$_2$Se$_3$ sample. Each curve represents a field cut at a different temperature (corresponding to Fig.~\ref{ip-specs}), with y-offset for clarity. The dotted line highlights uniform Kittel mode. At higher frequencies, additional resonance peaks corresponding to standing spin waves (SSWs) are visible, particularly at low temperatures. As temperature increases toward 25 K, the SSWs signatures become progressively weaker.}
    \label{ssw-Tdependency}
\end{figure}

%^%%%%%%%%%%%%%%%%%%%%%%%%%%%%%%%%%%%%%%%%%%%

\section{Resonant elastic X-ray scattering}{\label{REXS}}

\begin{figure}
    \centering
    \includegraphics[]{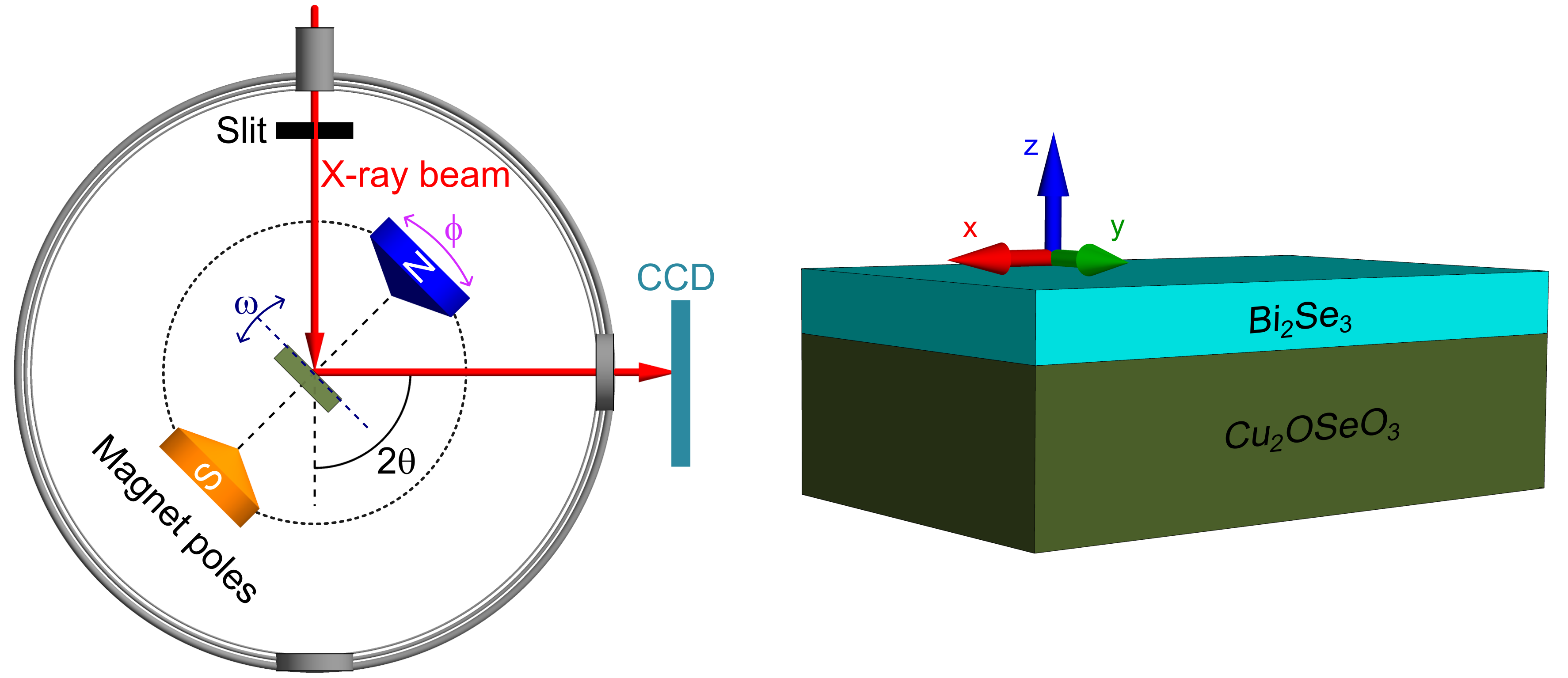}
    \caption{
    Schematic of the REXS setup in reflection geometry. The incident X-ray beam passes through a slit and impinges on the sample mounted at the center of a rotatable magnet assembly, allowing control of the magnetic-field orientation ($\phi$) and scattering angle ($2\theta$). The diffracted intensity is recorded by a CCD detector. 
    }

    \label{fig:Rexs_geo}
\end{figure}

\begin{figure}
    \centering
    \includegraphics[scale=0.95]{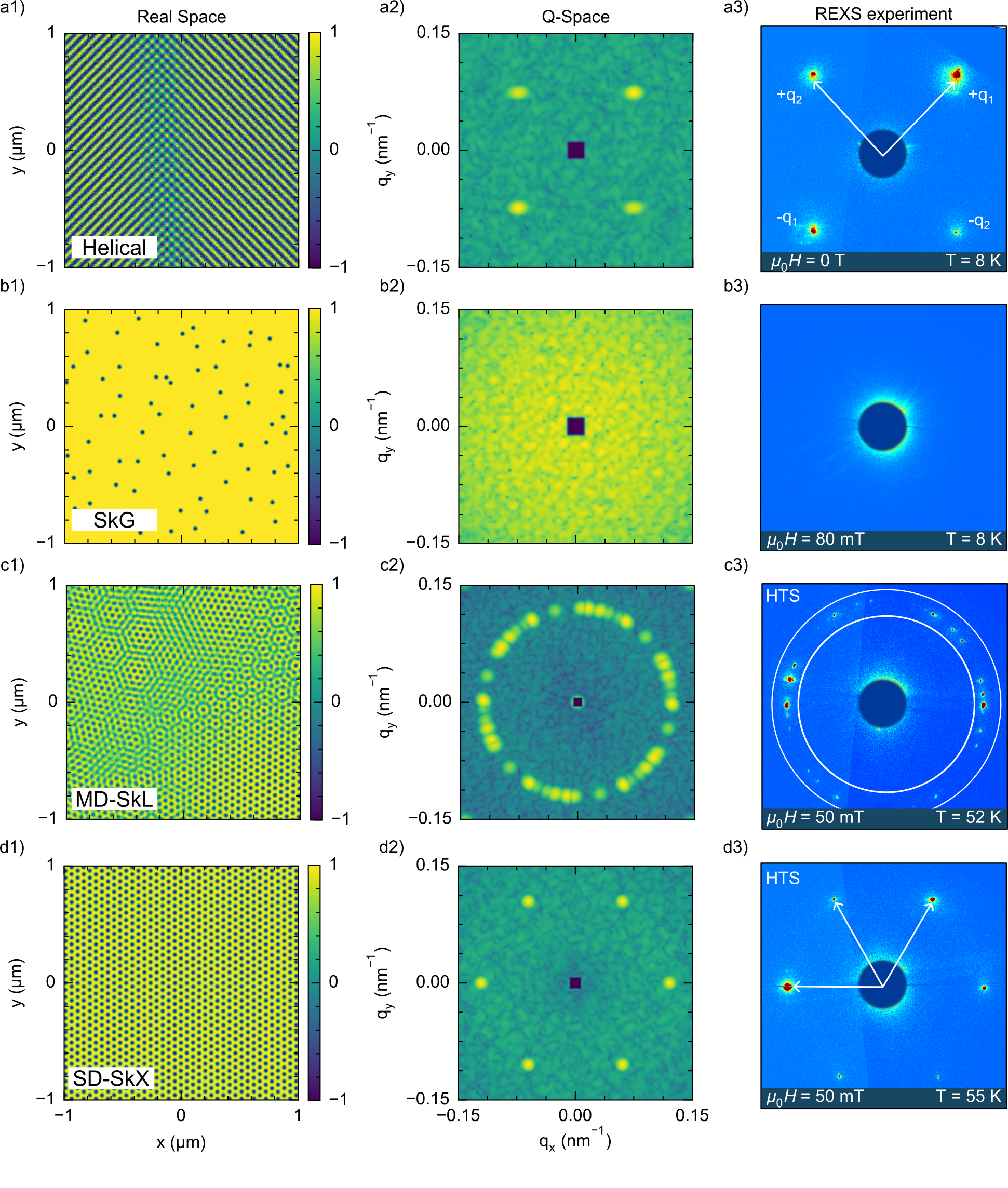}
\caption{Schematics s of real space magnetic configurations (left) with associated reciprocal space pattern (middle) compared with experimental:
(a1-a3) helical state with satellites at $\pm \mathbf{q}$; 
(b1-b3) skyrmion glass (SkG) lacking long-range order and magnetic satellites; 
(c1-c3) multi-domain skyrmion lattice (MD-SkL); 
(d1-d3) single-domain skyrmion lattice (SD-SkX) with six sharp Bragg peaks. 
The beamstop masks the direct beam at the center of Q-space images.
Synthetic REXS patterns were calculated from the Fourier intensity of the corresponding real-space $m_z(x,y)$ textures.
}

    \label{fig:REXS_all}
\end{figure}

% Resonant elastic X-ray scattering (REXS) measurements were carried out using the ALICE II endstation at the UE51$\_$PGM-1 OPUS undulator beamline, BESSY II (Helmholtz-Zentrum Berlin) in reflection geometry, as shown in Fig.~\ref{fig:Rexs_geo}. The incident soft X-rays were tuned to the Cu $L_3$ absorption edge (photon energy $\approx 932$~eV) to enhance magnetic contrast via resonant scattering from the Cu$^{2+}$ moments. The scattering angle was set to $2\theta \approx 96.5^\circ$, allowing detection of magnetic satellite peaks around the structural Bragg reflection. These magnetic satellites arise from the long-wavelength spin modulations and provide direct access to the helical, skyrmion (LTS and HTS), and TC magnetic phases in reciprocal space.
A schematic of the REXS geometry is shown in Fig.~\ref{fig:Rexs_geo}. The measurements were performed in reflection geometry with the magnetic field applied along ${\bf z} \parallel \langle 001\rangle$. In this configuration, magnetic satellite peaks around the structural Bragg reflection were used to identify the reciprocal-space fingerprints of the spin textures. The following section summarizes the correspondence between representative real-space textures, their synthetic reciprocal-space patterns, and the experimentally measured REXS images.
To compare the expected reciprocal-space signatures of different magnetic textures with the REXS data, we generated phenomenological two-dimensional real-space maps of the out-of-plane magnetization component $m_z(x,y)$. Skyrmion-lattice textures were represented by skyrmion-like cores arranged on a hexagonal lattice, while helical and conical textures were represented by sinusoidal single-$q$ modulations. Spatial coexistence of different textures was modeled by combining the corresponding real-space maps in different domains.
The synthetic reciprocal-space intensity was calculated from the two-dimensional Fourier transform of the real-space texture,
$$
I_{\mathrm{FFT}}(q_x,q_y)=
|
\mathrm{FFT}_{2D}
\big(
[m_z(x,y)-\langle m_z \rangle] W(x,y)
\big)
|^2 .
$$

Here, subtracting $\langle m_z \rangle$ removes the zero-$q$ component, and $W(x,y)$ denotes the numerical window/envelope function used to account for the finite real-space area and to suppress edge artifacts. The calculated intensity was normalized to its maximum value. These calculations are intended only to reproduce the qualitative reciprocal-space fingerprints of the expected magnetic textures and are not micromagnetic energy minimizations.

Figure~\ref{fig:REXS_all} provides a schematic overview of representative magnetic textures and their corresponding reciprocal-space scattering signatures, together with representative REXS measurements. The figure is intended to establish a direct correspondence between real-space magnetic order and the qualitative features observed in reciprocal space.

In the helical state Fig.~\ref{fig:REXS_all}(a), three symmetry-equivalent propagation vectors are allowed along the $\langle 100 \rangle$ directions in zero magnetic field. 
Due to the REXS geometry, the domain with an out-of-plane propagation vector is not detected, while the two orthogonal in-plane domains are observed. 
Accordingly, the reciprocal-space pattern exhibits four symmetric magnetic satellites at $\pm \mathbf{q}_1$ and $\pm \mathbf{q}_2$. 
The skyrmion glass (SkG) state \ref{fig:REXS_all}(b) consists of randomly distributed skyrmions without long-range positional order. Although individual skyrmions retain their internal spin structure, the lack of translational coherence means that no distinct magnetic satellites are observed in reciprocal space. This absence of sharp features reflects the absence of long-range periodic order, distinguishing the SkG state from the ordered skyrmion lattice phases.
In the multi-domain skyrmion lattice (MD-SkL) state \ref{fig:REXS_all}(c), multiple skyrmion lattice domains coexist within the illuminated region, each characterized by a distinct in-plane rotational orientation. In real space, this corresponds to a superposition of several hexagonal skyrmion lattices with different azimuthal alignments. In reciprocal space, each domain generates its own sixfold set of magnetic Bragg peaks. The resulting scattering pattern therefore exhibits a ring-like, sixfold-rotated peak set, whose relative intensities reflect the domain population within the probed area.
Finally, the single-domain skyrmion lattice (SD-SkX) state \ref{fig:REXS_all}(d) exhibits long-range hexagonal order with a unique lattice orientation. This single-domain configuration produces six sharp magnetic Bragg peaks arranged in a hexagonal pattern in reciprocal space. The well-defined peak positions and narrow widths indicate long-range coherence and a uniform lattice orientation across the probed region.

%%%%%
%moved
%%%%%

\begin{figure}
    \centering
    \includegraphics[]{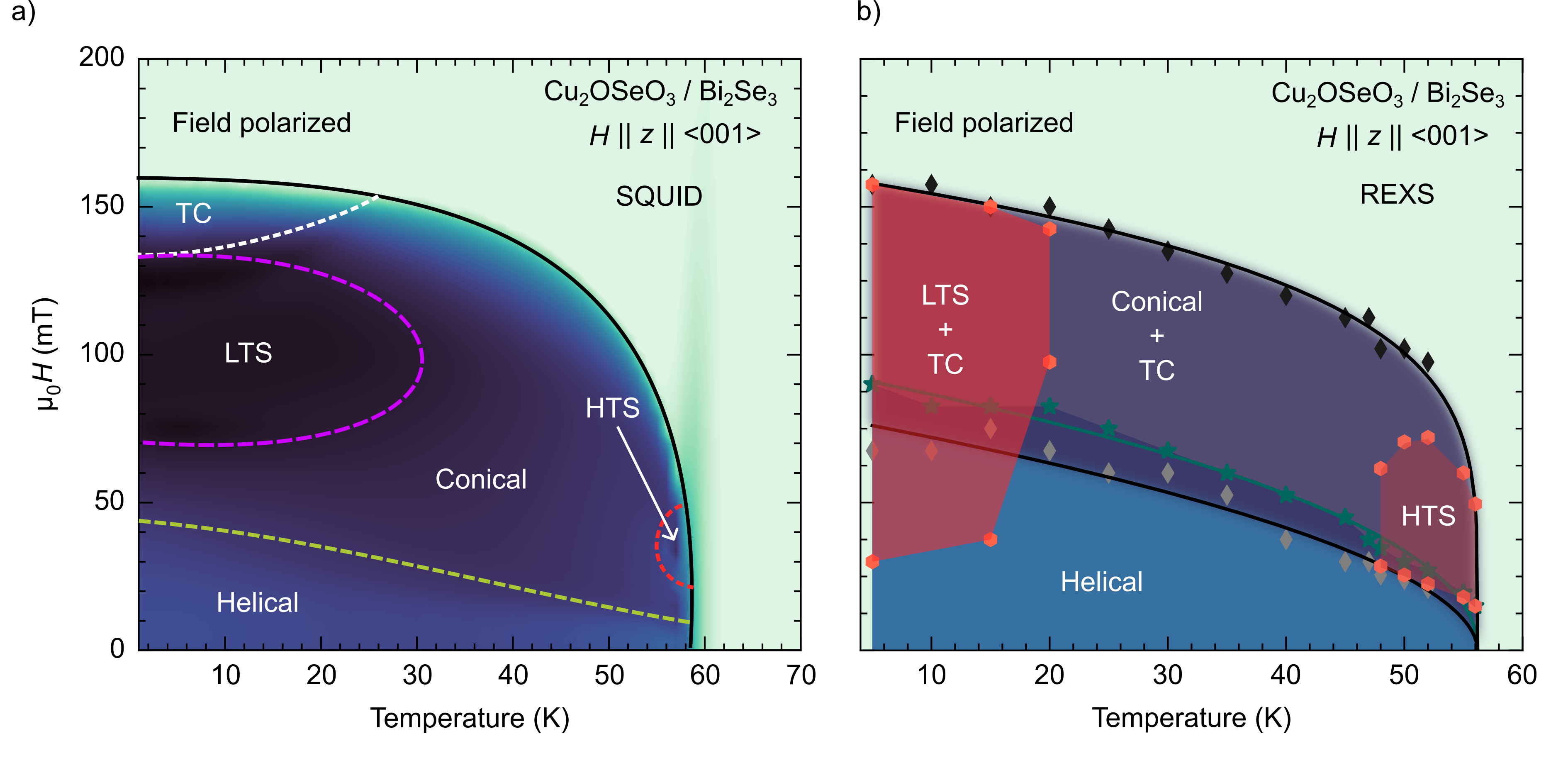}
    \caption{Magnetic phase diagrams of  Cu$_2$OSeO$_3$/Bi$_2$Se$_3$. (a) Bulk phase diagram derived from AC susceptibility measurements during magnetic-field sweeps from the FP state to zero. (b) Surface phase diagram obtained by mapping the phase boundaries from REXS measurements. The TC, LTS, and HTS refer to the tilted conical, low-temperature skyrmion, and high-temperature skyrmion phases, respectively.
    }
    \label{fig:SQUID_REXS_phase}
\end{figure}

Figures~\ref{fig:SQUID_REXS_phase} (a) and (b) present the magnetic phase diagrams of Cu$_2$OSeO$_3$/Bi$_2$Se$_3$ obtained from bulk AC susceptibility measurements and surface-sensitive REXS measurements, respectively. Although both measurements were performed on the same sample and in identical field geometry ($H \parallel {\bf z} \parallel \langle001\rangle$), clear differences are observed in the stability range and population of the low-temperature skyrmion phase.

In the REXS measurements, the LTS phase is not immediately stabilized under simple field sweeps. Instead, its characteristic six-fold diffraction pattern emerges only after applying a few magnetic-field cycles. This behavior reflects the metastable nature of the LTS phase and the presence of an energy barrier separating the conical and skyrmion states, as discussed before. Since REXS probes coherent long-range magnetic order at the surface, a well-defined diffraction pattern appears only once the skyrmion lattice is sufficiently populated and spatially ordered through repeated cycling.

In contrast, the SQUID AC susceptibility measurements reveal a clearly detectable LTS region in the bulk phase diagram even without an explicitly applied cycling protocol. We attribute this to the intrinsic field modulation of the AC susceptibility technique itself. During the measurement, the oscillating excitation field (333~Hz excitation frequency and 10~Oe modulation amplitude) continuously perturbs the magnetic state, effectively providing repeated minor field cycling around each set field value. This internal dynamic cycling can assist the system in overcoming the energy barrier between the conical and skyrmion states, thereby unintentionally populating the LTS phase in the bulk.

A direct comparison of the two phase diagrams (cf. \ref{fig:SQUID_REXS_phase}~a\&b) further reveals differences in the extent and boundaries of the LTS regime.
In the surface-sensitive REXS data, the LTS region spans a broader field range than in the bulk SQUID measurement, persisting to lower fields within the helical regime. This indicates an enhanced near-surface stability of the LTS phase relative to the bulk.

Moreover, the coexistence of TC and LTS phases detected by REXS highlights a modified near-surface magnetic environment that is not fully captured in bulk susceptibility measurements. These results suggest that the  Cu$_2$OSeO$_3$/Bi$_2$Se$_3$ interface modifies the local magnetic free-energy landscape. Interface-induced effects, such as altered magnetic anisotropy, strain, or proximity-driven interactions, can change the energy balance between the conical and skyrmion states. Consequently, the relative phase stability at the surface differs from that in the bulk, giving rise to a surface phase diagram that deviates from the bulk behavior.

Taken together, these results demonstrate that, despite identical sample and field geometry, bulk and surface probes reveal quantitatively different magnetic phase diagrams. The comparison provides direct evidence for interface-driven modifications of the magnetic free-energy landscape in the  Cu$_2$OSeO$_3$/Bi$_2$Se$_3$ heterostructure.

\section{Effective field and pitch change from the CCW splitting}
\label{app:field}

Across the skyrmion-field range the two CCW branches are separated by $\Delta f \approx 238$~MHz. Using the measured CCW field dispersion, $df/d\mu_0H \approx 7$~GHz/T, this maps to an effective-field difference between the bulk and interfacial skyrmion environments,
\begin{equation}\label{eq:Heff}
  \mu_0\,\Delta H_{\mathrm{eff}} = \frac{\Delta f}{\,df/d\mu_0H\,}
  \approx 34~\mathrm{mT}.
\end{equation}
The Zeeman-equivalent value $\Delta f/(\gamma/2\pi)\approx 8.5$~mT, with $\gamma/2\pi = 28$~GHz/T, applies only to a Larmor-like mode; the gyrotropic CCW mode disperses $\sim$4 times more weakly with field, so Eq.~\eqref{eq:Heff} is the appropriate conversion. Attributing the difference to an enhanced interfacial DMI, the chiral energy scale obeys $\mu_0 H_{c2}\propto D^2/A$ and the helimagnetic wavevector $q\propto D/A$; since the skyrmion mode frequency scales as $f\propto H_{c2}\propto D^2$,
\begin{equation}\label{eq:dD}
  \frac{\Delta D}{D} \approx \tfrac{1}{2}\,\frac{\Delta f}{f}\approx 5\%,
  \qquad
  \frac{\Delta q}{q} \approx \frac{\Delta D}{D}\approx 5\%,
\end{equation}
the factor $\tfrac12$ reflecting $f\propto D^2$ but $q\propto D$. The interfacial skyrmion lattice is thus expected to have a propagation vector $\sim$5\% larger (pitch shortened by $\sim$3~nm from $\sim$60~nm). As this is comparable to the magnetic-satellite linewidth, the bulk and interfacial populations appear as a single broadened LTS satellite rather than a resolved doublet.

%%%%%%%%%%%%%%%%%%%%%%%%%%%%%%%%%%%%%%%%%%%%%%%%%%%%%%%%%%%%%%%%%
%%%%%%%%%%%%%%%%%%%%%%%%%%%%%%%%%%%%%%%%%%%%%%%%%%%%%%%%%%%%%%%%%
%%%%%%%%%%%%%%%%%%%%%%%%%%%%%%%%%%%%%%%%%%%%%%%%%%%%%%%%%%%%%%%%%

\section{Fitting model of the resonance spectra}\label{app_fit}

To analyze the FMR spectra of the  Cu$_2$OSeO$_3$/Bi$_2$Se$_3$, we performed frequency-domain fitting of the field derivative of the background-corrected transmission parameter \( S_{21} \) for each fixed external magnetic field. The VNA measures the transmission coefficient \( S_{21} = ( {V_{\text{2}}}/{V_{\text{1}}}) e^{i\phi} \), where \( V_{\text{1}} \) and \( V_{\text{2}} \) are the input and output voltages at ports 1 and 2, respectively, and \( \phi \) represents the electrical length of the setup. In the FMR experiment using CPW, a voltage proportional to the high-frequency susceptibility is induced in the CPW by the precessing magnetization.
The fitting model was based on a central difference approximation of the magnetic susceptibility \cite{maier2018note}, expressed as

\begin{equation}
 d_D S_{21} = i \omega A e^{-i\phi} \frac{\chi(\omega + \Delta \omega_+) - \chi(\omega - \Delta \omega_+)}{2 \Delta \omega_+}, 
\end{equation}

where \( A \) is the complex amplitude that incorporates modulation sensitivity, \( \phi \) is a phase offset accounting for lineshape asymmetry. The frequency modulation step is defined as \( \Delta \omega_+ = \Delta H_+ \left( \partial \omega / \partial H_{\text{ext}} \right) \approx \Delta H_+ \gamma \mu_0 \), assuming a linear dependence of frequency on magnetic field. 

The dynamic susceptibility \( \chi \) was modeled using a high-frequency Lorentzian form:

\begin{equation}
    \chi(\omega, H_{\text{ext}}) = \frac{\omega_M (\gamma \mu_0 H_{\text{ext}} + i \Delta \omega)}{(\omega_{\text{res}}(H_{\text{ext}}))^2 - \omega^2 + i \omega \Delta \omega},
\end{equation}

where \( \omega_{\text{res}} \) is the resonance frequency, \( \omega_M = \gamma \mu_0 M_s \) corresponds to the saturation magnetization, \( \Delta \omega \) is the full linewidth at half maximum, and \( \gamma \) is the gyromagnetic ratio. 

\begin{figure}[ht]
    \centering
    \includegraphics[]{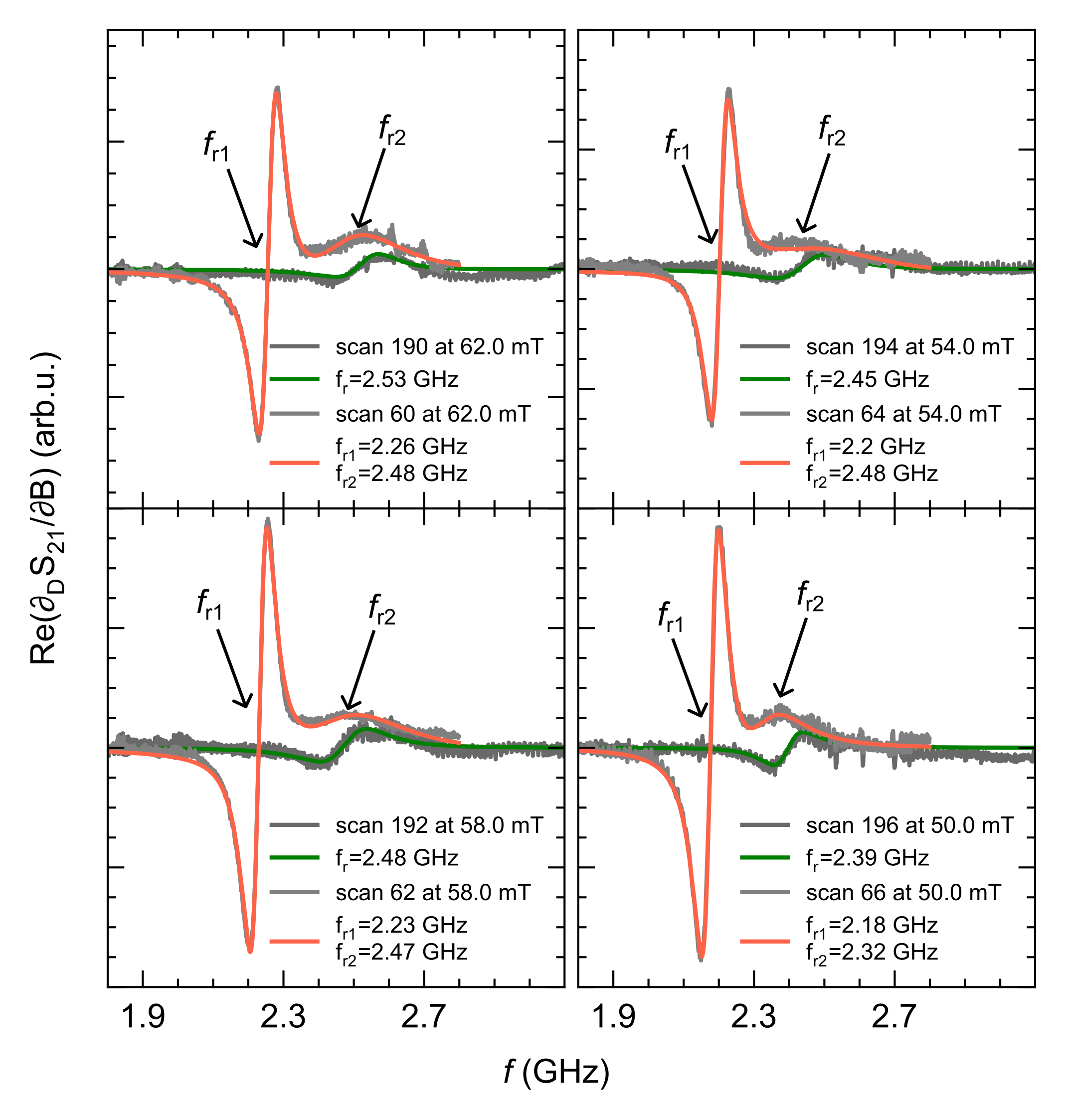}
    \caption{Comparison of line scans of the resonance spectra: scans with (orange) and without (green) magnetic field cycling. Each panel presents two representative scans at the same field, one from the cycling protocol and one from the non-cycled case, highlighting two distinct modes. The fitted resonance frequencies \( f_\text{r} \) are labeled for clarity. These scans correspond to the skyrmion mode region discussed in the main text (see Fig.~\ref{all_plot})
}
    \label{fourplots}
\end{figure}

Figure~\ref{fourplots} shows fitting of the field cuts at four different magnetic fields using the derivative divide method corresponding to the experimental spectra of Fig.~\ref{spec22} in the main manuscript.

\end{document}